\documentclass[aps, pra, reprint,groupedaddress]{revtex4-1}

\usepackage{graphicx}
\graphicspath{ {images/} }

\usepackage[usenames, dvipsnames]{color}

\usepackage{amsmath}

\begin{document}


\title{Observations of a Field-Aligned Ion/Ion-Beam Instability in a Magnetized Laboratory Plasma}

\author{P. V. Heuer}
\email[]{pheuer@physics.ucla.edu}

\author{M. S. Weidl}

\author{R. S. Dorst}

\author{D.  B. Schaeffer}
\altaffiliation{Princeton University}

\author{A. S. Bondarenko}

\author{S. K. P. Tripathi}

\author{B. Van Compernolle}

\author{S. Vincena}

\author{C. G. Constantin}

\author{C. Niemann}
\affiliation{University of California, Los Angeles}

\author{D. Winske}
\affiliation{Los Alamos National Laboratory}


\date{\today}

\begin{abstract}

Collisionless coupling between super Alfv\'{e}nic ions and an ambient plasma parallel to a background magnetic field is mediated by a set of electromagnetic ion/ion-beam instabilities including the resonant right hand instability (RHI). To study this coupling and its role in parallel shock formation, a new experimental configuration at the University of California, Los Angeles utilizes high-energy and high-repetition-rate lasers to create a super-Alfv\'{e}nic field-aligned debris plasma within an ambient plasma in the Large Plasma Device (LAPD). We used a time-resolved fluorescence monochromator and an array of Langmuir probes to characterize the laser plasma velocity distribution and density. The debris ions were observed to be sufficiently super-Alfv\'{e}nic and dense to excite the RHI. Measurements with magnetic flux probes exhibited a right-hand circularly polarized frequency chirp consistent with the excitation of the RHI near the laser target. We compared measurements to 2D hybrid simulations of the experiment.

\end{abstract}


\maketitle

\section{Introduction}
\label{introduction}

Collisionless shocks are generated by the interaction of a super-Alfv\'{e}nic beam plasma and a magnetized background plasma. The characteristics of a collisionless shock are strongly dependent on the angle $\theta$ of the shock normal to the background magnetic field. Shocks where $\theta \approx 0^\circ$ are termed parallel shocks. Parallel collisionless shocks are common phenomena in space and astrophysical plasmas, and have been studied theoretically since the early 1960's~\cite{Parker1961, Sagdeev1966}. Within the solar system, \textit{in-situ} measurements of parallel and perpendicular shocks have been made by spacecraft crossing through naturally occuring structures such as planetary bow shocks~\cite{Treumann2009, Mazelle2003, Burgess2005quasi} and comets~\cite{Omidi1994}. Spacecraft shock crossings provide the most direct measurements of shocks in nature, but inherently provide a spatially and temporally limited data set. Some other shock systems are not accessible to current spacecraft such as supernovae remnants~\cite{Ostriker1988, Spicer1990}. Laboratory experiments such as the ones described in this paper can therefore complement spacecraft measurements by producing comprehenisve data sets under controlled conditions to validate theory and simulation~\cite{Drake2000}. 

A number of unique features make parallel and quasi-parallel shocks a subject of particular interest. Spacecraft have observed density and temperature anomalies within parallel shocks whose origin is not yet understood~\cite{Thomas1991hybrid, Parks2017shocks}. Simulations show that parallel shocks (at high Mach numbers~\cite{Omidi1990low}) are not steady-state, but continuously reform~\cite{Burgess1989cyclic} in a turbulent process that has never been observed in the laboratory. Theory and simulations have also shown that parallel shocks are capable of efficiently accelerating particles through the process of first order Fermi acceleration~\cite{Blandford1987particle}. This process is capable of accelerating particles to the very high energies (up to $10^{19}$ eV) observed in the cosmic ray spectrum~\cite{Bell2004turbulent}. First order Fermi acceleration has never been directly observed in nature or in the laboratory.

Shocks where $\theta \approx 90^\circ$ are termed perpendicular shocks. In a perpendicular shock, Larmor coupling transfers energy from a beam (or ``piston'') plasma to the background plasma~\cite{Papadopoulos1987, Bondarenko2017}, leading to the development of a spatially confined shock in the ambient plasma with a characteristic length scale on the order of one ion-inertial length~\cite{Newbury1998ramp}. This process has previously been replicated in laboratory experiments with dimensionless parameters scaled to astrophysical phenomena\cite{Niemann2014, Schaeffer2017, Schaeffer2017generation}.

In the parallel direction, the beam plasma couples to the background plasma through a set of electromagnetic ion beam instabilities~\cite{Winske1984, Gary1991}. The properties of these beam instabilities determines the structure of the resulting shock, so understanding these instabilities is essential to understanding parallel shocks. The beam instabilities are classified in terms of the beam densities and velocities that excite them, and by the polarization (sense of rotation in time at a fixed position) and helicity (sense of rotation in space at fixed time) of the electromagnetic waves they drive. The polarization of a wave with real frequency $\omega_r$ can be defined as~\cite{Gary1991}:
\begin{equation}
\label{def_pol}
P = \pm \frac{\omega_r}{| \omega_r |}
\end{equation}
Where $+1$ and $-1$ correspond to right-hand and left-hand circular polarization respectively. The helicity of a wave with wave number $k$ propagating parallel to the background field ($k \times B_0 = 0$) can be defined as~\cite{Gary1991}:
\begin{equation}
\label{def_hel}
\sigma = \pm \frac{k}{  | k | }
\end{equation}
Where $+1$ and $-1$ are described as positive and negative helicity respectively. 

Two instabilities of primary importance are the right-hand resonant instability (RHI) and the non-resonant instability (NRI). Both instabilities grow more quickly at higher beam densities and velocities, however the dominant instability changes as the beam density and velocity are increased. The NRI is a non-cyclotron-resonant instability dominant for high beam densities and highly super-Alfv\'{e}nic beam velocities~\cite{Winske1984}. Waves produced by the NRI have negative helicity and are left-hand circularly polarized in the laboratory frame. The RHI  is a cyclotron resonant mode dominant for lower beam densities relative to the background density and low (but still super-Alfv\'{e}nic) beam velocities relative to those required for the NRI~\cite{Winske1984}. The RHI is right-hand circularly polarized with positive helicity in the laboratory frame of reference. Two other instabilities, the left-hand resonant instability and the ion cyclotron anisotropy instability, are only competitive for very high or anisotropic temperature beam plasmas, respectively~\cite{Gary1985, Winske1985}. 

In the process of coupling, these instabilities launch magnetosonic waves and shear Alfv\'{e}n waves. Along with particles leaked through or reflected by the shock front, these waves generate a turbulent region (or ``foreshock'') upstream of the shock~\cite{Burgess1995, Burgess2005quasi}. The resulting shock is spatially much larger than in the perpendicular case, forming over hundreds of ion-inertial lengths~\cite{Weidl2016}. Parallel shocks also develop more slowly than perpendicular shocks (time scales of $\tau_\parallel \sim 100$ $\Omega_{c}^{-1}$ compared to $\tau_\perp \sim \Omega_{c}^{-1}$), limited by the growth rate of the coupling instabilities.  

The formation of parallel shocks has been simulated by both 1D and 2D hybrid codes, which support the conclusion that the RHI and NRI play the dominant roles in shock formation~\cite{Winske1984, Gary1991, Weidl2016}. Simulated beam instabilities grow exponentially, driving large amplitude fluctuations in the magnetic field which eventually saturate~\cite{Winske1984, Gary1991}. Recent 2D hybrid simulations~\cite{Weidl2016} show that it is feasible to observe the beginning stages of parallel shock formation within the length scales of the Large Plasma Device (LAPD)~\cite{Gekelman2016} at the University of California, Los Angeles (UCLA).

A new experimental configuration at UCLA combines two high-energy lasers~\cite{Niemann2012} and the LAPD to study the RHI and its role in parallel shock formation. A super-Alfv\'{e}nic laser-produced plasma ``beam'' streaming anti-parallel to the background magnetic field, was created within the ambient plasma of the LAPD. The densities and velocities of the laser produced plasma were measured with Langmuir probes and a time-resolved fluorescence monochrometer. Right-hand circularly polarized waves in the transverse magnetic field were observed with fast, dense beams, but not with slower, more tenuous beams. These results are consistent with excitation of the RHI in a region near the laser target. Hybrid simulations under similar conditions qualitatively reproduce the observed waves, but predict significantly faster instability growth rates.

This article is organized as follows. Section~\ref{theoretical_background} defines notation and introduces a linear model of the RHI. Details of the experimental setup are discussed in Section~\ref{experimental_setup}. Analysis of the data is presented in Section~\ref{analysis}, and the conclusions are summarized in Section~\ref{conclusion}.

\section{Theoretical Background}
\label{theoretical_background}

\begin{figure}[b]
\centering
\includegraphics[width = 0.5\textwidth]{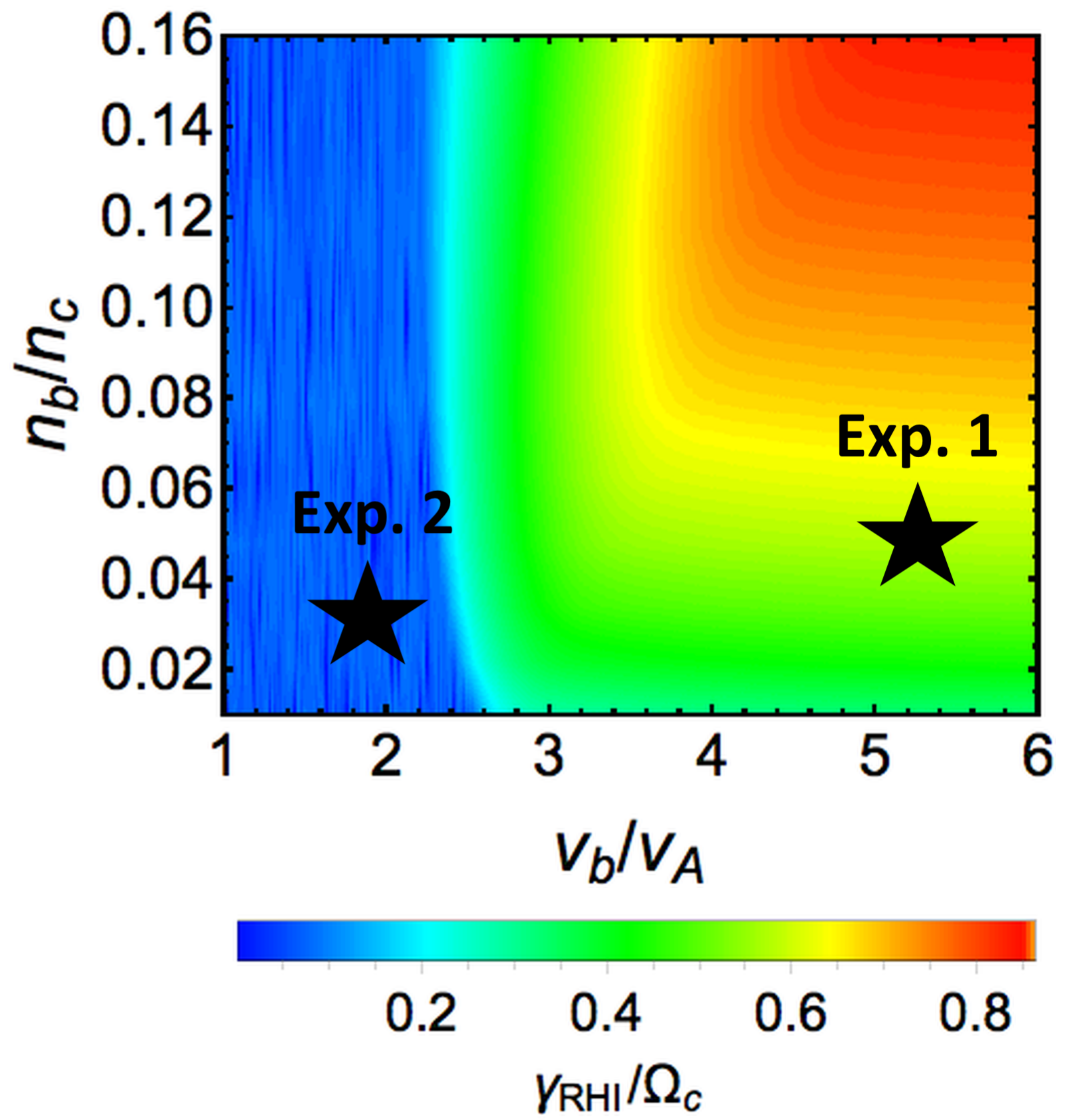}
\caption{\label{rhi_growth}Maximum growth rate of the RHI in the lab frame derived from cold-plasma Vlasov theory (Eq.~\ref{disp_rel}) over a range of beam densities and velocities for a beam of $C^{+4}$ ions. Two stars indicate the locations of both experiments in this parameter space. The plotted points correspond to the fastest debris observed in each experiment.}
\end{figure}

The interacting plasma beams are modeled as a charge-neutral three-component plasma consisting of debris (or ``beam'') ions, background (or ``core'') ions, and electrons, denoted by subscripts $b$, $c$, and $e$. The cyclotron and plasma frequencies for each species $s$ are defined as  $ \Omega_s = q_s B / m_s c$ and $\omega_{p,s} = \sqrt{4 \pi n_s  Z_s^2 e^2 / m_s} $. $B_0$ denotes the background magnetic field. In keeping with previous literature, we adopt the following conventions. Frequencies are normalized to the background ion cyclotron frequency, velocities to the Alfv\'{e}n velocity $V_A = B (4 \pi n_c m_c)^{-\frac{1}{2}}$, masses to the background ion mass $m_c$, and charges to the background ion charge $q_c$. Number densities are normalized to the total density so that $n_c + q_b n_b =  n_e = 1$, and wavenumbers are scaled to the background plasma ion inertial length $\delta_i = c / \omega_{pi}$.  The Alfv\'{e}nic Mach number is defined as $M_A = v_b / v_A$. We define $k > 0$. Primes denote quantities in the stationary-electron frame, while those without primes are in the laboratory frame.

A dispersion relation for the RHI generated by interacting bi-Maxwellian beams can be derived from linear Vlasov theory~\cite{Gary1978, Gary1984}. Applying the inertia-less electron limit ($\Omega_e \gg \omega' - k v_e'$) and assuming that the beams are temperature-isotropic leads to the following simplified normalized dispersion relation (for $k \times B_0 = 0$ and $q_c = 1$) in the stationary-electron frame~\cite{Weidl2017towards}:
\begin{equation}
\begin{aligned}
0 = &\bigg(\frac{{\omega'}^2}{\omega_{p,ref}^2 } - k^2\bigg) (\omega' - k v_c' + \Omega_c)(\omega' - k v_b' + \Omega_b)\\
&-  \frac{n_b q_b^2}{m_b} (\omega' - k v_b')(\omega' - k v_c' + \Omega_c)
\\ &- n_c  (\omega' - k v_c') (\omega' - k v_b' + \Omega_b)
\\ &+ n_e \frac{\omega' - k v_e'}{\Omega_c}(\omega' - k v_c' + \Omega_c)(\omega' - k v_b' + \Omega_b)
\end{aligned}
\label{disp_rel}
\end{equation}
Where $\omega_{p,ref}^2 = 4\pi n_e e^2 / m_c$ Additionally, assuming the plasma is current-free in the stationary-electron frame:
\begin{equation}
v_c' = - \frac{q_b}{q_c} \frac{n_b}{1 - q_b n_b} v_b'
\label{v_c}
\end{equation}
The resulting quartic equation is then solved for the growth rate $\gamma = \operatorname{Im}(\omega)$ of the RHI over a range of beam parameters (Fig.~\ref{rhi_growth}). Growth increases with both beam velocity and density. In the high-Mach number limit $v_b' >> v_A$, the growth rate for the RHI can be approximated by~\cite{Winske1986, Gary1991}: 
\begin{equation}
\gamma_{RHI} \propto \bigg (   \frac{n_b}{n_e} \bigg )^{1/3}
\label{rhi_growth_eq}
\end{equation}
The theoretical RHI growth rate slowly asymptotes to zero as $n_b \rightarrow 0$ and $M_A \rightarrow 1$. In practice, a higher velocity threshold exists where the RHI ceases to remain competitive with other modes and will no longer be observed. The behavior of the RHI in this low Mach number limit has previously been the subject of simulations of parallel shock formation~\cite{Omidi1990low}.

\section{Experimental Setup}
\label{experimental_setup}

\begin{figure}
\centering
\includegraphics[width = 0.48\textwidth]{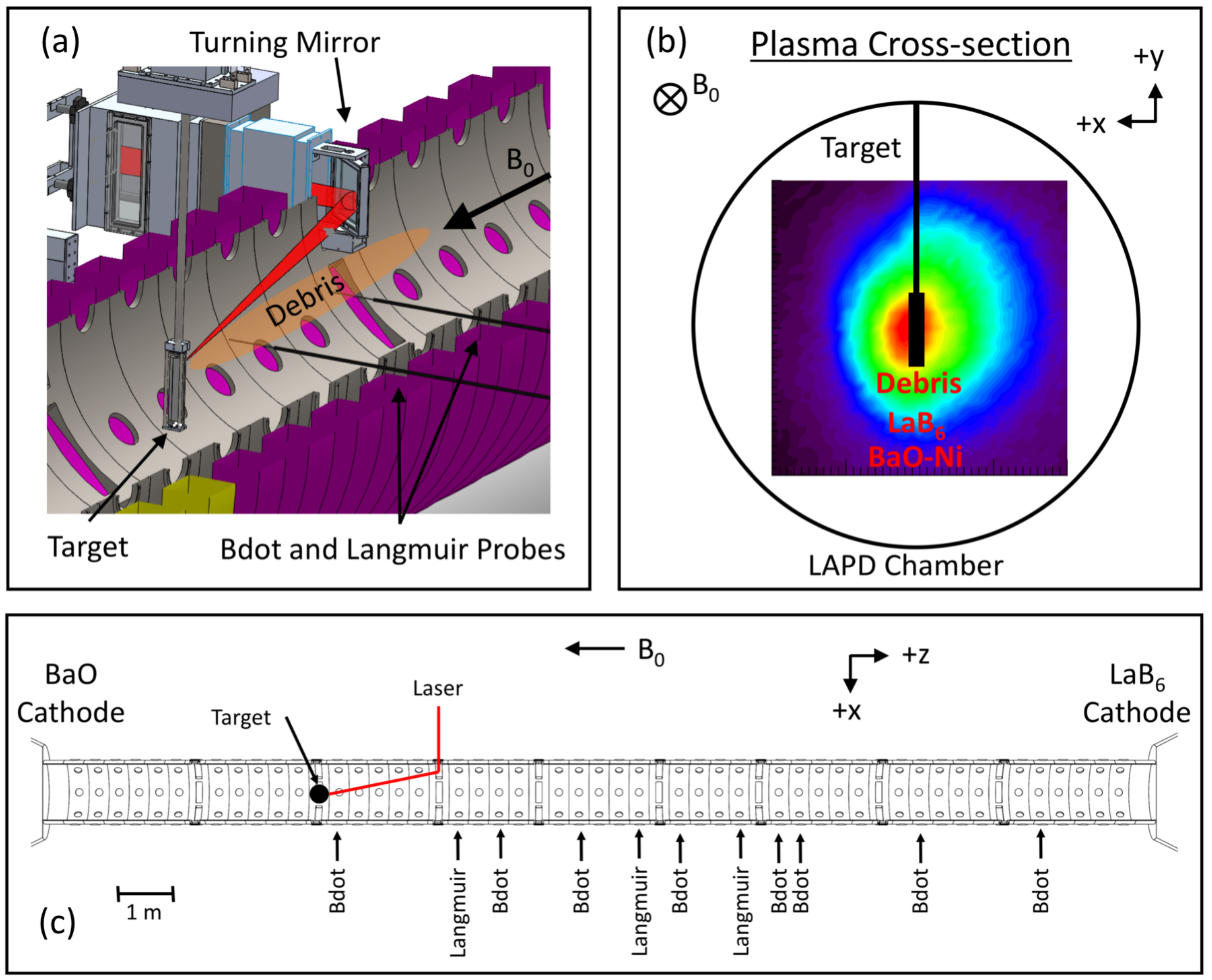}
\caption{\label{exp_setup} Experimental setup. a) The laser was directed onto the target through a steering mirror inside the vacuum chamber. Debris ions were ablated anti-parallel to the background magnetic field. b) The target (and therefore the debris ions) were centered on the dense LAPD LaB$_6$-generated helium background plasma. c) A top-view of the LAPD shows the laser's path into the chamber, and an array of magnetic flux `bdot' probes and Langmuir probes used to diagnose the debris-background interaction.} 
\end{figure}

\begin{table}
\begin{tabular}{ l |  c  |   c  }
Parameters & Exp.~1 & Exp.~2\\
\hline
Laser System & \textbf{Raptor} & \textbf{Peening} \\
Energy (J) &  \textbf{200} & \textbf{15} \\
Intensity (W/cm$^2$) & $10^{11}$  & $10^{11}$ \\
Shots/hr &  \textbf{1} & \textbf{3600} \\
Target Material & \textbf{C$\mathbf{_2}$H$\mathbf{_4}$} &\textbf{Graphite}\\
Ambient Plasma & He & He \\
$B_0$ (G) & $300$ & $300$ \\
$n_c$  (cm$^{-3}$) & $9 \times 10^{12}$ & $9 \times 10^{12}$\\
$T_{c,i}$ (eV) & $1$ & $1$\\
$T_{c,e}$ (eV) & $5$ & $5$\\
\hline
Measured Values & & \\
\hline
Max. $v_b$ (km/s) & $\mathbf{550}$ & $\mathbf{200}$ \\
$v_b$ at max. $n_b$ (km/s) & $\mathbf{300}$ & $\mathbf{100}$ \\
Max. $n_b$ (cm$^{-3}$) & $\mathbf{5\times 10^{11}}$ & $\mathbf{8 \times 10^{11}}$\\
$n_b$ at max. $v_b$ (cm$^{-3}$) & $\mathbf{1\times 10^{11}}$ & $\mathbf{2 \times 10^{11}}$\\
\end{tabular}
\caption{Representative parameters for both experiments, with differences in bold. $B_0$ is the ambient background field and $n_c$ the ambient plasma density. $T_{c,i}$ and $T_{c,e}$ are the ambient plasma ion and electron temperatures respectively.}
\label{exp_param}
\end{table}

\begin{table}
\begin{tabular}{ l |  c }
Species & Line Wavelength(s) (nm) \\
\hline
$C^{+ 1}$ (CII)  & 426.752 \\
$C^{+ 2}$ (CIII)  & 229.687 \\
$C^{+ 4}$ (CV) & 227.089, 494.399 \\
\end{tabular}
\caption{Carbon species observed by time-resolved monochromatic ion fluorescence measurements.}
\label{carbon_wavelengths}
\end{table}

Two experiments were conducted under similar conditions and with similar diagnostics, but using two separate laser systems that are part of the UCLA High Energy Density Plasma (HEDP) Phoenix Laser Laboratory~\cite{Niemann2012}. Experiment 1, conducted using the high-energy Raptor laser (1053 nm, 25 ns, 200 J, 1 shot/hr), was designed to maximize the debris velocity and density. Due to the limited shot rate, study of the instabilities was largely confined to one dimension, along the debris plasma blowoff axis. Experiment 2 used the high-repetition rate (but lower energy) Peening laser~\cite{Hackel1993}(1053 nm, 15 ns, 15 J, 3600 shots/hr) to collect lineouts and 2D planes with good statistics (3-5 shots/position) and spatial resolution (5-10 mm step sizes). 

Both lasers were focused through an \textit{f}/34 lens to the same intensity of $I \approx 10^{11}$ W/cm$^2$ onto a target centered in the ambient plasma (Fig.~\ref{exp_setup}a,c). After focusing, the beam was deflected by a steering mirror inside the vacuum chamber (protected by an anti-reflective coated glass blast shield) onto the target. Since the debris velocity scales with laser intensity~\cite{Grun1981characteristics, Schaeffer2016} both lasers produced similar debris velocities despite the order-of-magnitude difference in laser energy. Both cylindrical and planar targets were used, made out of either high density polyethylene (HDPE, $C_2 H_4$) or graphite (Table~\ref{exp_param}). The target was rotated and translated after each laser shot to present a fresh target surface. Although the laser beam met the target surface at an angle, the debris expanded parallel to the surface normal and the background magnetic field. 

The ambient plasma was created in the LAPD~\cite{Gekelman2016}, operated by the Basic Plasma Science Facility (BaPSF). The LAPD is a cylindrical plasma device with a solenoidal magnetic field. The LAPD plasma volume is 20 m long and 1 m in diameter with a variable axial magnetic field (0.2-1.8 kG) and gas fill (H, He, Ne, etc.). The plasma was ionized by two concentric cathodes at either end of the cylindrical chamber. A BaO (barium oxide) coated cathode created a low-density plasma ($n \approx 2 \times 10^{12}$ cm$^{-3}$) 60 cm in diameter, while a LaB$_6$ (Lanthanum Hexaboride) cathode created a smaller, higher density plasma ($n \approx 10^{13}$ cm$^{-3}$) 40 cm in diameter. The ambient plasma was highly repeatable, steady state (for 10 ms), quiescent, and net current free. The ion and electron temperatures were $T_i = 1$ eV and $T_e = 5$ eV respectively. The LAPD experimental parameters were identical for both experiments (Table~\ref{exp_param}). At these parameters, the background ion cyclotron frequency was $\Omega_{c,i} = 7 \times 10^{5} $ rad/s, the ion inertial length $\delta_i = $ 15 cm,  the Alfv\'{e}n velocity $v_A \approx 100$ km/s, and the ratio of the background plasma pressure to the background magnetic field pressure $\beta = 10^{-3} $.  Dimensionless variables throughout this paper are scaled to these parameters.

We define the following coordinate system in the LAPD for convenience (Fig.~\ref{exp_setup}c). The axial background magnetic field defines the $-\hat z$ axis, and the target surface lies in the x-y plane. The point where the laser meets the target is defined as the spatial origin \{x,y,z\} = \{0,0,0\}, and the laser fires at time t=0. 

Magnetic field measurements were made using an array of 3-axis differentially wound magnetic flux ``bdot'' probes, 3 mm in diameter~\cite{Everson2009}. The induced current was passed through a 100 MHz differential amplifier before being digitized. The signal was then integrated to measure changes in the vector magnetic field. The ion saturation current (``isat'')  was measured by an array of 1.32 mm$^2$ area planar Langmuir probes facing towards the laser target. Langmuir probes were negatively biased to -150 V with respect to the chamber walls to repel electrons. The isat was converted to a voltage by a 9.5 $\Omega$ resistor, which was then digitized after passing through an optical isolator. All signals were digitized by a 1.25 GHz, 10-bit digitizer. 

Probes were initially aligned visually along the z axis of the LAPD using a calibrated surveyor's transit (accuracy $\approx \pm$ 2 mm), then positioned either by motorized probe drives~\cite{Gekelman2016} (resolution $\approx \pm$  0.5 mm) or by hand (resolution $\approx \pm$ 2 mm) between laser shots. Exp.~2 utilized motorized probe drives to automatically collect large rectangular planes of data ($\approx$ 1600 positions $\times$ 3-5 repetitions) with a spatial grid spacing of 0.5 cm in both the XZ and XY planes. During Exp.~1, probes were moved both by motorized drives and by hand, taking care to preserve the orientation of the probe relative to the background field. At the length scales of the features observed ($ >1$ cm), the only advantage of motorized drives for Exp.~1 was convenience. 

Debris charge states were observed independently by time-resolved monochromatic ion fluorescence measurements at different wavelengths (Table~\ref{carbon_wavelengths}). Ion fluorescence was collected by a fiber probe near the target. The fiber probe collected light over a $90^\circ$ angle through a 75 mm focal length lens inserted into the plasma. The light was imaged onto a linear array of $40 \times 100$  $\mu$m UV-grade glass fibers, which were coupled to a narrow band ($< 1$ nm) monochromator (1800 lines/mm), then collected by an avalanche photodiode (APD). This resolution was sufficient to distinguish emission lines from different debris charge states. The fiber probe lens was moved between 20-30 cm from the z axis based on the target wavelength (Table~\ref{carbon_wavelengths}) in order to correctly focus the measured light. 

We define circular polarization relative to the ambient magnetic field. Viewed parallel to the ambient magnetic field, a right-hand polarized field vector rotates clockwise, and a left-hand polarized field vector counter-clockwise. In both experiments the debris streamed \textit{anti}-parallel to the ambient B-field, while simulations (Section~\ref{simulations}) were conducted (in keeping with previous literature) with debris streaming parallel to the field. Analysis of wave polarization therefore depends on the orientation of the field and debris.

\section{Analysis}
\label{analysis}

\subsection{Debris Characterization}
\label{debris_characterization}

\begin{figure*}
\centering
\includegraphics[width = 1\textwidth]{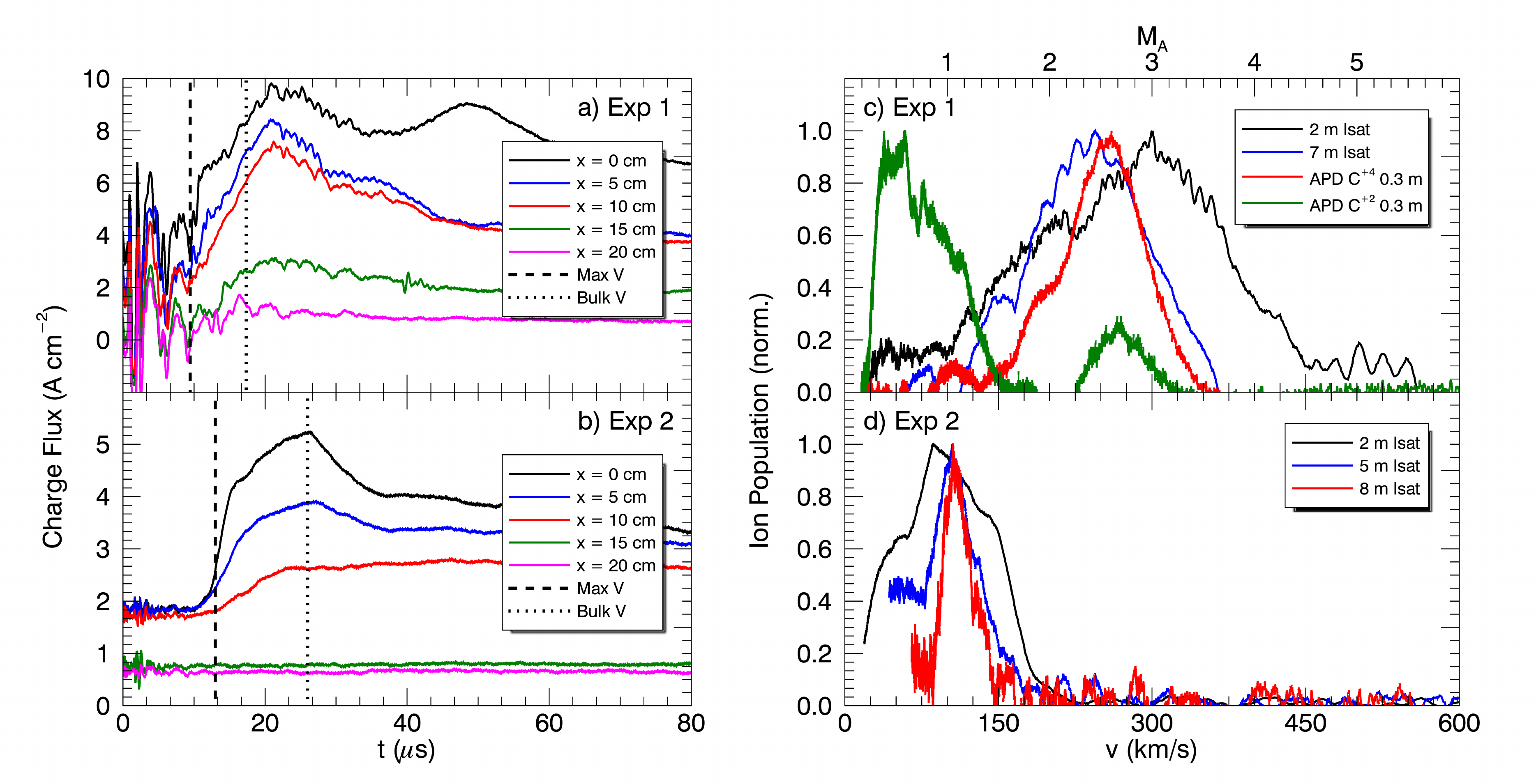}
\caption{\label{lang}  a,b)  Ion charge flux, derived from isat, measured at different x positions. Measurements were made at $z=5.2$ m and $z=2.6$ m during experiments 1 and 2 respectively. Density peaked on the z axis then fell off with increasing x. Oscillations early in time were caused by fast laser-produced electrons~\cite{Niemann2013}. Density decreased with x more quickly in Exp.~2 than Exp.~1. c,d). Time-of-flight debris velocity distributions (normalized and inherently integrated over all charge species) were measured by Langmuir probes at various distances from the target during both experiments. Velocity distributions of specific charge states measured by time-resolved monochromatry using an APD near the target are also plotted for Exp.~1. The Alfv\'{e}nic Mach number $M_A$ was calculated with respect to the background Alfv\'{e}n speed of $\sim 100$ km/s. }
\end{figure*}

The laser produced plasma ablated from the target consists of different species of carbon ions, as well as protons and electrons. Previous experimental measurements and simulations with the 1-D radiation hydrodynamics code HELIOS show that higher charge states move with substantially higher velocities, and that C$^{+4}$ is the most common charge state produced at these experimental conditions~\cite{Schaeffer2016}. The debris has a cos$^2 (\phi)$ initial angular velocity distribution~\cite{Heuer2016}, where $\phi$ is the angle away from the target surface normal. Therefore, only a fraction of the debris is truly field-aligned. The resulting debris profile is densest and has the highest field-parallel velocity on axis, which implies that the maximum RHI growth rate is also on axis (Eq.~\ref{rhi_growth_eq}). Measurements conducted during both experiments confirmed that the beam density peaks on axis (Fig.~\ref{lang}a,b). 

Time-resolved ion fluorescence measurements collected at a known distance from the target through a monochromator (``monochromatry'') were used to measure the velocity distribution of the most populated charge states by time-of-flight. Ion fluorescence cannot easily be used to measure density~\cite{Heuer2016}, and is too dim to detect far from the target. Therefore, these measurements are well complemented by isat measurements from several down-field Langmuir probes (facing towards the target). Langmuir probes provide an estimate of charge-integrated density. Isat measurements were also used to estimate the charge state integrated ion velocity distribution by time-of-flight farther from the target, where ion fluorescence signals were dim. 

Monochromatry was only performed during Exp.~1. The predominant bulk charge state was confirmed to be C$^{+4}$ by comparison of monochromatry with Langmuir probe measurements of the bulk density. Fig.~\ref{lang}c shows that the peak of the C$^{+4}$ velocity distribution agrees well with the peak of the measured isat. Two separate C$^{+4}$ transition lines were observed with good agreement. C$^{+2}$ ions were also observed, but these ions travel more slowly than the C$^{+4}$ ions, corresponding to a low velocity tail on the charge-integrated velocity distribution (Fig.~\ref{lang}a). Time-of-flight velocity distribution estimates for both of the observed charge states are consistent with simulation predictions~\cite{Schaeffer2016}.  Neither of the observed velocity distributions of carbon charge-states  extend to the high velocities seen by the Langmuir probes. This suggests that the high velocity tail consists of higher charge states of carbon (C$^{+5}$, C$^{+6}$) or protons.

Isat measurements taken by Langmuir probes at known distances from the target were also used to estimate the charge-state-integrated debris ion velocity distribution by time-of-flight (Fig.~\ref{lang}c,d). Two important velocities measured were the velocity at peak density (``bulk velocity'') and the highest velocity ions observed (``maximum velocity"). Exp.~1 produced a bulk velocity of 300 km/s ($M_A \approx 3$) with a maximum velocity of 550 km/s ($M_A \approx 5$) (Fig.~\ref{lang}a,c). Exp.~2 produced a bulk velocity of 100 km/s ($M_A \approx 1$) with a maximum velocity of 200 km/s ($M_A \approx 2$) (Fig.~\ref{lang}b,d). 

Isat measurements were also used to estimate the density of the debris plasma. Neglecting sheath effects, we approximate the density measured by a probe of circular area $A$ collecting all particles of velocity $v$ and charge $q$ in a cylindrical volume $V = Avt$ over a time $t$ as:
\begin{equation}
n_c = \frac{1}{q}\frac{It}{Avt} = \frac{J}{qv}
\label{langmuir_n_eq}
\end{equation}
where $I$ and $J$ are the current and current density respectively. Isat measurements were made during both experiments. Two important densities measured were the maximum density and the density of the highest velocity ions, both measured on the LAPD z axis ($x=y=0$). In Exp.~1, the maximum density measured was $5\times10^{11}$ cm$^{-3}$ ($0.05$ $n_e$), and the density of the highest velocity ions was $1\times10^{11}$ cm$^{-3}$ ($0.01$ $n_e$), both measured at $z=5.2$ m from the target (Fig.~\ref{lang} c). In Exp.~2, the maximum density measured was $8\times10^{11}$ cm$^{-3}$ ($0.08$ $n_e$), and the density of the highest velocity ions was $2 \times 10^{11}$ cm$^{-3}$ ($0.02$ $n_e$), measured at $z=2.0$ m from the target (Fig.~\ref{lang}d). The higher density measured during Exp.~2 may be due to the closer position of the probe to the target.

Previous experimental work has shown that the velocity of mass ablated by a laser from a planar target scales with the absorbed laser intensity and the laser wavelength, but not the incident laser energy~\cite{Grun1981characteristics, Meyer1984experimental, Schaeffer2016}. Since the laser intensity and wavelength were identical between Exp.~1 and Exp.~2, these scaling laws do not explain the difference in velocities observed. One possible explanation is that the presence of protons from the $C_2 H_4$ target increased the velocities observed during Exp.~1.

The theoretical RHI growth rate was calculated based on the measured beam densities and velocities, shown as stars in Fig.~\ref{rhi_growth}. Their location in parameter space suggest that the RHI should be observable in Exp.~1, but should not grow appreciably in Exp.~2.

\begin{table}
\begin{tabular}{ l |  c }
Loss Mechanism  & Time Scale ($\mu$s) \\
\hline
Collisions $\tau_{i,i}$ & $4 \times 10^4$ \\
Collisions $\tau_{i,e}$ & 200 \\
Collisions $\tau_{e,i}$ & \textbf{0.05} \\
Collisions $\tau_{e,e}$ & \textbf{0.1}\\
C$^{+4}$ Recombination & $8\times 10^4$ \\
C$^{+4} \to$ C$^{+3}$ Charge Exchange & $1 \times 10^3$\\
\end{tabular}
\caption{Mean free path time scales for various density loss mechanisms calculated for conditions matching these experiments. $\tau_{b,c}$ denotes the collision time scale between a species of the beam plasma $b$ and a species of the core plasma $c$. The recombination rate includes both radiative and collisional three-body recombination. Time scales shorter than the relevant experimental time scales of $\sim 10$ $\mu$s are highlighted in bold.}
\label{loss_times}
\end{table}

\begin{figure}
\centering
\includegraphics[width = 0.5\textwidth]{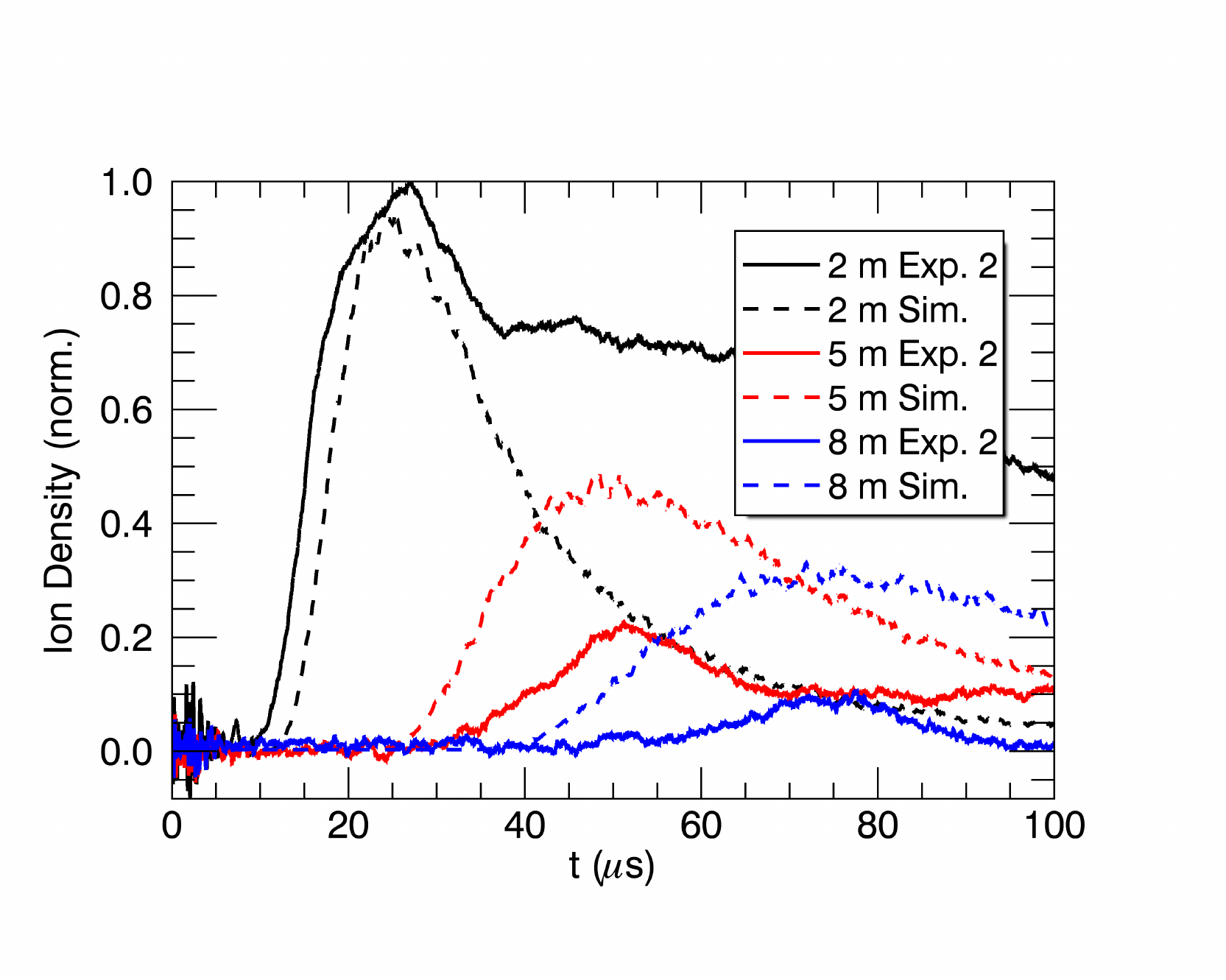}
\caption{\label{monte_carlo_density} A 1D Monte-Carlo simulation compared to measured densities inferred from isat during Exp.~2. 1D longitudinal velocity dispersion explains some but not all of the density loss observed. All simulated measurements were normalized to the 2 m simulated trace, and all experimental measurements were normalized to the 2 m measured trace. The simulated density decreases on further probes due to velocity dispersion. The measured densities decreased more than predicted by the simulation, indicating that velocity dispersion cannot account for all of the density loss.}
\end{figure}

Growth of the RHI decreases substantially when either the debris density or velocity falls below the instability threshold. Comparison of time-of-flight measurements between Langmuir probes at different z positions (Fig.~\ref{lang}a,b) indicates that the bulk debris ions did not decelerate. In contrast, the beam density was measured to decrease substantially along the length of the experiment (making the small population of high-velocity ions difficult to detect with Langmuir probes). Understanding and reducing this density loss is crucial to promoting sustained growth of the RHI and NRI.

Atomic and collisional processes resulting in scattering or charge neutralization cannot explain this decrease in density: the experiment time scales are shorter than the characteristic time scales for radiative and three-body recombination~\cite{Rumsby1974, Nahar1997} and charge exchange~\cite{Dijkkamp1985} (Table~\ref{loss_times}). The beam ions are collisionless with the background plasma. Although beam electrons are collisional with both background ions and electrons~\cite{NRLformulary}(Table~\ref{loss_times}), the background electrons maintain quasi-neutrality so that scattering of the beam electrons does not affect the beam ions. 

Part of the decrease in density can be attributed to longitudinal velocity dispersion of the measured velocity distribution (Fig.~\ref{lang}a,b). As the distribution propagates, the spacing between particles of different velocities increases. This appears as a decrease in density on the Langmuir probes, which measure density in discrete time intervals (Eq.~\ref{langmuir_n_eq}). A simple 1-D Monte-Carlo simulation (Fig.~\ref{monte_carlo_density}) was conducted to determine how much density loss can be attributed to this effect. A sample population of simulated ions were initialized with a 1D velocity distribution matching observations from the Langmuir probe at 2 m during Exp.~2. The ions were then time evolved, assuming that the their velocities remain constant. The ion density was measured by three virtual Langmuir probes at positions corresponding to probe locations from Exp.~2. The results show that the measured density decreases about twice as fast as the simulated density, suggesting that velocity dispersion explains about half of the density loss observed. The remaining decrease is likely caused by transverse expansion, and will be investigated in future 2D and 3D simulations.

\subsection{Frequency and Polarization Analysis of Magnetic Field Oscillations}
\label{magnetic_field_waves}

\begin{figure}
\centering
\includegraphics[width = 0.5\textwidth]{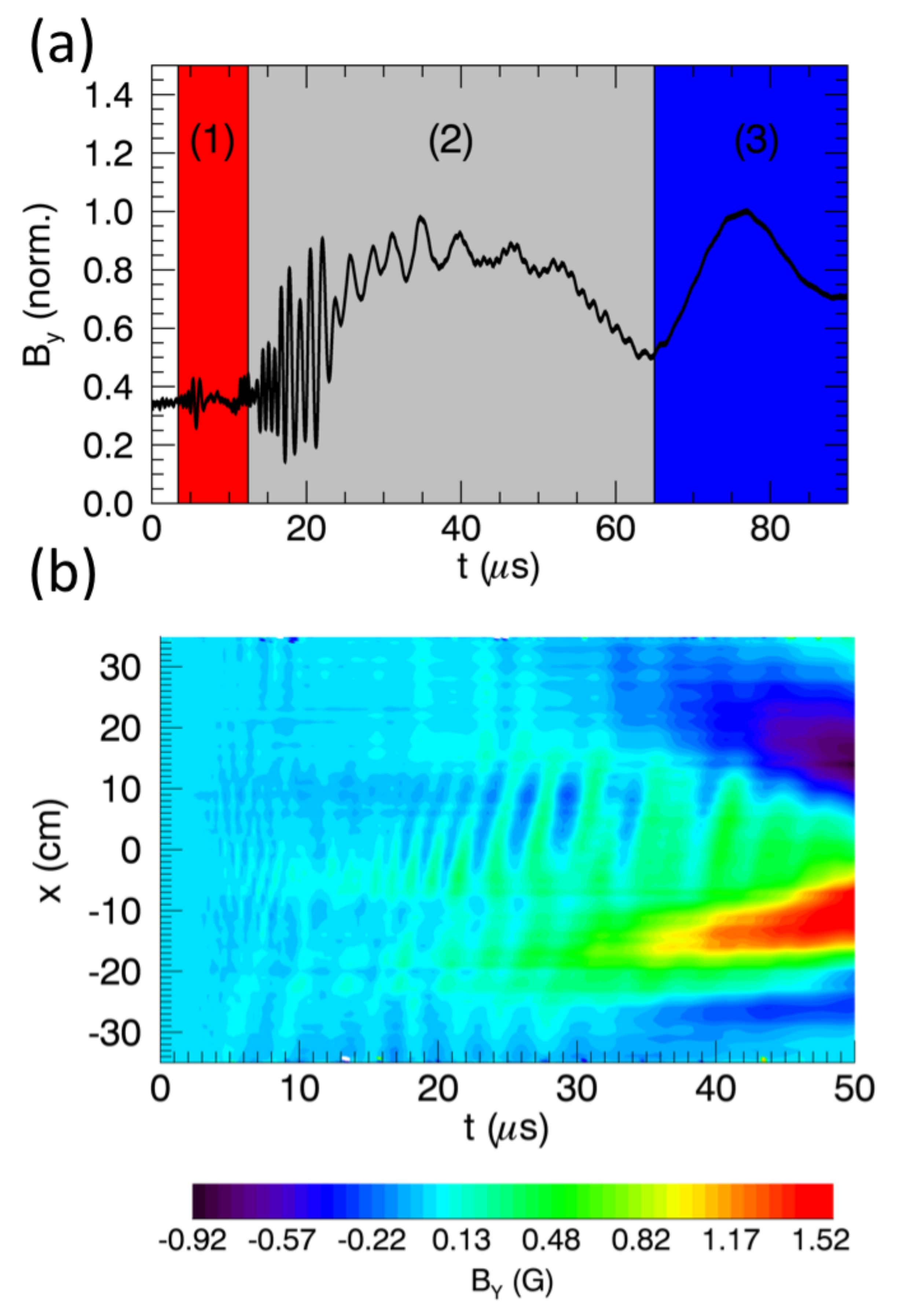}
\caption{\label{waves_combined} Three wave regions separated by velocity are observed in the transverse magnetic field. (a) A magnetic flux probe at $\{x,y,z\} = \{0,0,7.5$ m$\}$  during Exp.~1  shows three separate wave regions. A high frequency, low amplitude wave packet (1) is followed first by a larger amplitude high frequency chirp (2) and then a shear Alfv\'{e}n wave (3). (b) An x lineout at $z=7.5$ m from the target taken during Exp.~2 shows the arrival of high-frequency plane waves ($ \delta B \approx 0.25$ Gauss), followed by a higher amplitude shear Alfv\'{e}n wave centered on the z axis. }
\end{figure}

\begin{figure}
\centering
\includegraphics[width = 0.5\textwidth]{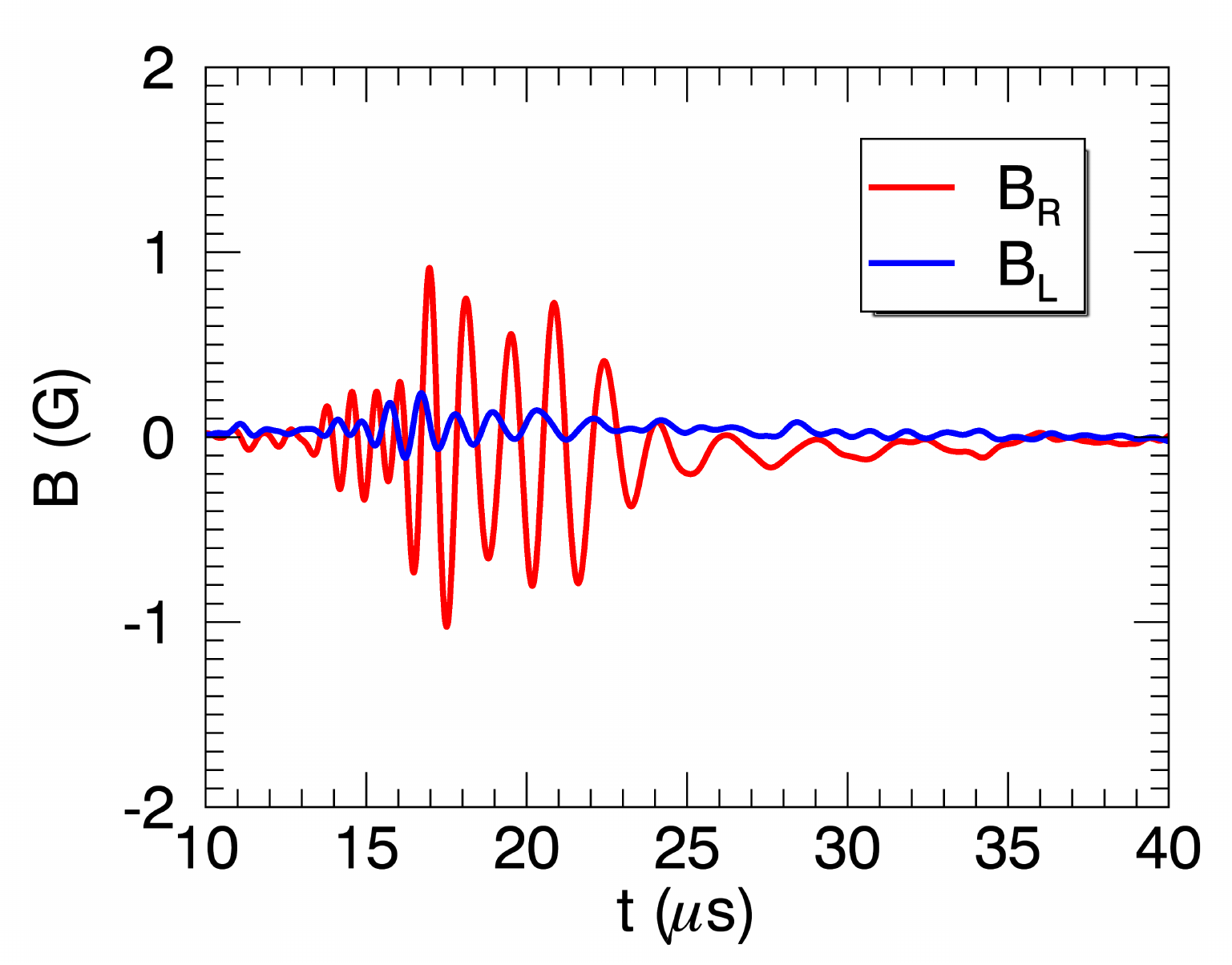}
\caption{\label{b_pol}Decomposition of a sample magnetic field trace from Exp.~1 at  $\{x,y,z\} = \{0,0,7.5$ m$\}$ onto a circularly polarized basis. The dominance of the oscillations ($\sim 1$ MHz) in $B_R$ demonstrates that the wave is right-hand circularly polarized. Signals have been filtered in frequency space around 1 MHz to remove a low-frequency offset.}
\end{figure}

\begin{figure}
\centering
\includegraphics[width = 0.5\textwidth]{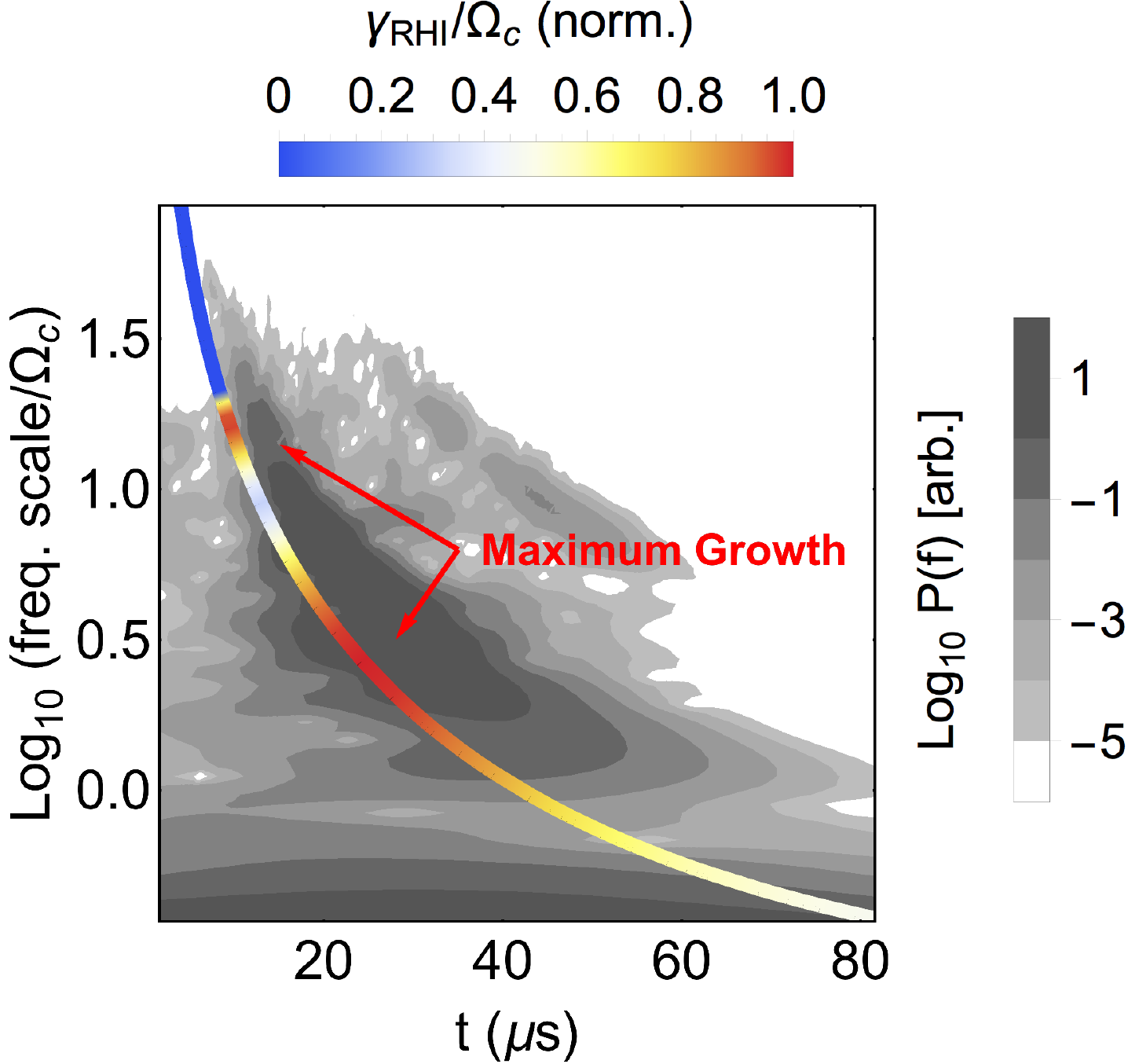}
\caption{\label{disp_plot} A frequency power spectrum $P(f)$ of transverse magnetic field waves observed during Exp.~1 was calculated using a Morlet-wavelet transform (greyscale contours). Darker contours indicate more power. A time-of-arrival fit based on a linear cold-plasma linear model (Eq.~\ref{disp_rel}) is overlayed on the data. The color of the fit line denotes the calculated RHI growth rate $\gamma_{RHI} (\omega)$ normalized to the background ion cyclotron frequency $\Omega_c$.}
\end{figure}

Waves were observed in the transverse magnetic field by magnetic flux probes during both experiments. The frequency range of the waves was determined using a Morlet wavelet transform. The polarization of each wave was determined by decomposing the measured transverse magnetic field components ($B_X$ and $B_Y$) onto circularly polarized basis vectors. The right-handed and left-handed components of the magnetic fields ($B_R$ and $B_L$) were calculated by shifting the relative phases of the signals, then superimposing them in Fourier space~\cite{Terasawa1986decay, Weidl2016}:
\begin{equation}
\widetilde{B}_{L/R} = \frac{1}{2}( \widetilde{B}_X \pm i \widetilde{B}_Y)
\end{equation}

A comparison of transverse magnetic field data from arrays of magnetic flux probes shows three distinct wave features in both experiments (Fig.~\ref{waves_combined}a).  For the purpose of discussion, we will refer to them (in ascending order of time of arrival) as: 1) the lower hybrid wave, 2) the chirp, and 3) the shear Alfv\'{e}n wave. These features appear on both transverse components ($B_X$ and $B_Y$). We discuss the lower hybrid wave and shear Alfv\'{e}n wave first, before a more detailed analysis of the chirp.

The lower hybrid wave is a high frequency ($\omega \approx 1$ MHz) wave packet which filled the entire cross-sectional area of the LAPD plasma (Fig.~\ref{waves_combined}b, 5 - 30 $\mu$s) at a low amplitude ($0.1$ G) in both experiments. These waves arrived slightly earlier ($t< 1$ $\mu$s) at negative x positions than positive x positions, which may be due to a density gradient along the x axis.  The waves were linearly polarized at an angle of $\approx 45^\circ$ below the x axis. The early arrival time of these waves cooresponds to a velocity of 1000 - 1600 km/s ($M_A \approx 10 - 16$), which suggests that they were created by particles moving at those speeds. This range of velocities is consistent with the laser-produced electron beam~\cite{Niemann2013}. Previous experiments on the LAPD~\cite{Vincena2008quasielectrostatic} have shown that fast laser-produced electrons can drive quasi-electrostatic whistler waves (lower hybrid waves) near and above the background plasma lower hybrid frequency, $\omega_{LH}^2 = \Omega_{e}\Omega_{i} / (1 + \Omega_{e}^2 / \omega_{p,e}^2)$, where $\omega_{p,e}$ is the electron plasma oscillation frequency. At these experimental parameters (Table~\ref{exp_param}), $\omega_{LH} = 1.5$ MHz. The approximate match between $\omega_{LH}$ and the frequency of the wave observed in region 1 (Fig.~\ref{waves_combined}a) suggests that this wave is a lower hybrid wave~\cite{Vincena2008quasielectrostatic}.

The low frequency ($\omega < \Omega_c$) shear Alfv\'{e}n wave propagated at the measured Alfv\'{e}n speed of $\sim 100$ km/s in both experiments. Shear Alfv\'{e}n waves have been previously studied in the LAPD~\cite{VanZeeland2001, VanCompernolle2008, Gekelman2011, Tripathi2015}. In our experiments the wave continued to grow in amplitude as it propagated, reaching maxima slightly off axis (x = 5 cm) of $\delta B / B_0 = 0.13$ ($\delta B = 40$ G) in Exp 1 and $\delta B / B_0 = 0.03$ ($\delta B = 8$ G) in Exp 2. The polarity of the shear wave ($+ \hat y$ for negative x, $- \hat y$ for positive x, Fig.~\ref{waves_combined}) is consistent by Ampere's Law with a beam of positive ions moving in the negative z direction. 

The chirp began at $\omega \approx 10$ $\Omega_c$ then approached $\omega = \Omega_c$ over a period of $30$ $\mu$s. In both experiments the leading edge of the chirp arrived at the same time as the fastest particles (Fig.~\ref{waves_combined}).  Waves reached a maximum amplitude slightly off axis ($x = 5$ cm) of $\delta B / B_0 = 0.03$ ($\delta B = 10$ G) during Exp.~1 and $\delta B / B_0 = 0.006$ ($\delta B = 2$ G) during Exp.~2. Projecting the magnetic fields onto a circularly polarized basis shows that the high-frequency waves observed in Exp.~1 were right-hand circularly polarized (Fig.~\ref{b_pol}). In Exp.~2, right hand circularly polarized waves were only observed near the target ($z<$5 m). Beyond 5 m, waves in Exp.~2 became linearly or elliptically polarized.

A Morlet wavelet transform was used to produce a power spectrum of the chirp (Fig.~\ref{disp_plot}, greyscale contours). The main chirp, starting at $15$ $\mu$s, contained the majority of the wave energy. A second weaker chirp observed starting at $30$ $\mu$s was a reflection of the first chirp from the LAPD LaB$_6$ anode (the point of reflection was determined by comparing the time-of-arrival of a single frequency in each chirp on multiple magnetic field probes). Similar chirps were observed in both experiments.

A theoretical prediction for the growth rate of the waves and their earliest arrival time at a probe was made using the RHI dispersion relation (Eq.~\ref{disp_rel}) for a single beam ion charge state and velocity. All waves were assumed to originate at $z=0$, which is a reasonable approximation because the beam density and therefore the growth rate was highest near the target. The waves then traveled through the plasma, dispersing by phase velocity to the probes downfield (following Eq.~\ref{disp_rel} applied to just the background plasma, i.e. $v_b = n_b = 0$). In the current experimental regime, the theoretical growth rate can be either singly or doubly peaked as a function of frequency, determined by the beam density and velocity. The actual growth rate is a convolution of several charge states each with a non-zero-width velocity distribution.  

Figure~\ref{disp_plot} compares a theoretical growth rate calculated to the observed wave power spectrum. The growth rate was calculated for the most populated ion species, C$^{+4}$, with a density of $n_b = 0.08$ $n_e$ and a speed of $M_A \approx 4$ based on the measurements presented in Section~\ref{debris_characterization}. The linear model agreed well with both the arrival time of each measured frequency and the range over which the instability grows. The theoretical curve was calculated for waves originating at the target at $t=0$, and therefore overlaps the leading edge of the observed chirp. The most highly excited frequency scales (between $0$ and $1.25$) align with a maximum in the observed power spectra. We postulate that a better fit could be achieved by convolving the theoretical growth rates for the full range of ion species and velocities measured in the experiment, weighted by their relative densities. Carrying out this calculation requires more knowledge of the debris charge state distribution, and is therefore left for future work.

\subsection{Comparison to Simulation}
\label{simulations}

\begin{figure}
\centering
\includegraphics[width = 0.46\textwidth]{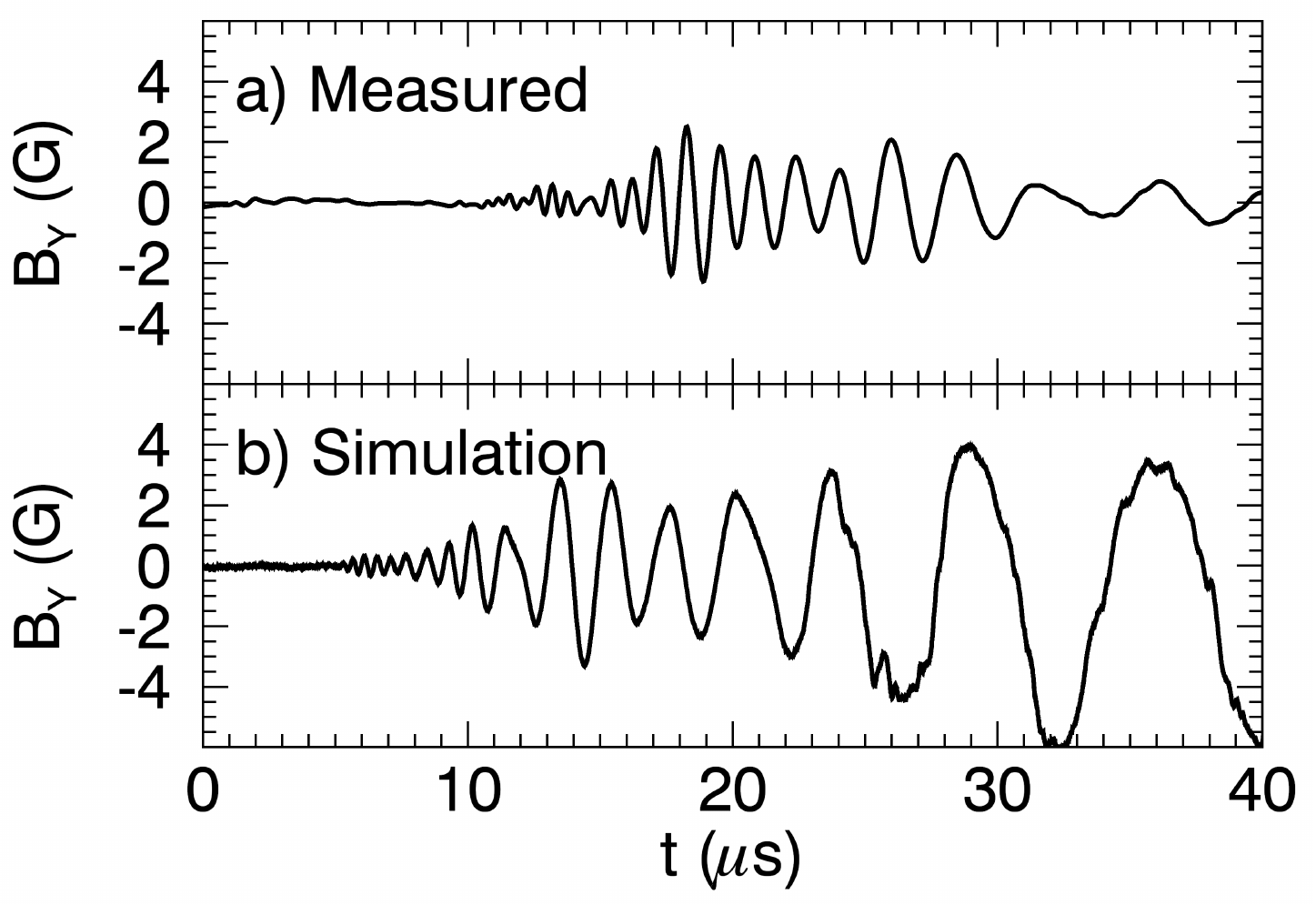}
\caption{\label{sim_vs_data} (a) Experimentally measured $B_Y$ from Exp.~1. (b) Simulated $B_Y$ measurement matching experimental conditions for Exp.~1.  Both traces exhibit a similar frequency chirp over a similar frequency range ($10$ $\Omega_c$ to $\Omega_c$). The measured signal has been filtered in frequency space around 1 MHz to remove a low-frequency offset.}
\end{figure}

Simulations of each experiment were conducted using a hybrid code~\cite{Weidl2016, Weidl2017towards} that represented ions with a particle-in-cell (PIC) technique and electrons as an MHD fluid. Two carbon debris species ($C^{+3}$ and $C^{+4}$) were initialized with a velocity distribution matching the experimental measurements. A background of He$^{+1}$ was loaded with uniform density throughout the simulation domain, which was several times longer than the LAPD. Fields were computed and stored on a grid of cells with side length $\delta_i/4$ with periodic boundary conditions on the longitudinal boundaries (along the debris axis), but lossy transverse boundaries to represent debris particle losses to the LAPD walls. It was found that lossy particle boundary effects must be included in order for the amplitude of waves in the simulated magnetic field to match experimental observations.

Figure~\ref{sim_vs_data}b shows that the simulated beam ions produced waves in the transverse magnetic field as they propagated. The waves formed a chirp over a frequency range of $10$ $\Omega_c$ to $\Omega_c$, which agrees well with experiments. A power spectrum produced by a wavelet analysis of the simulated magnetic field shows that the arrival times of individual frequencies were comparable to the experimental power spectrum shown in Figure~\ref{disp_plot}. However, there are several important differences between the simulated and measured magnetic field traces shown in Figure~\ref{sim_vs_data}. The amplitude early in time ($t<15$ $\mu$s) was much greater in the simulated signal than the experiment, suggesting that higher frequencies modes grew faster in the simulation. Around $t=20$ $\mu$s, the amplitude of the measured waves was comparable to the simulation. Later in time ($t > 25$ $\mu$s), the amplitude of the measured signal again decreased while the simulated field continued to grow.  This discrepancy could be explained by a drop in particle density in the experiment at that time that was not reflected in the simulation, which is a subject of continuing research.

\section{Conclusion}
\label{conclusion}

The right-hand resonant instability (RHI) plays an important role in the formation of collisionless parallel shocks in space and astrophysical plasmas. The RHI is a cyclotron-resonant instability between a tenuous super Alfv\'{e}nic beam plasma and a denser background plasma, which drives right-hand polarized magnetosonic waves in the background plasma.

A new series of experiments combined the high-energy Raptor laser (Exp.~1) and the high repetition rate Peening laser (Exp.~2) with the LAPD at UCLA to study the formation of this instability. Three transverse magnetic field wave features were observed in both experiments. At early time, a packet of low amplitude, high frequency, linearly polarized oscillations was observed with a velocity corresponding to fast electrons or protons created at the target. A higher amplitude chirp from $\approx 10$ $ \Omega_c$ to $\Omega_c$ immediately followed at a velocity corresponding to the fastest debris ions in the experiment. The final feature observed was a shear Alfv\'{e}n wave.

The chirp produced in Exp.~1 was right-hand circularly polarized throughout the experiment, consistent with the theoretical description of the RHI. The otherwise similar chirp in Exp.~2 was right hand circularly polarized near the target, but linear or elliptically polarized at distances greater than $5$ m from the target. These observations agree with linear theory for measured beam ion densities and velocities, which shows that the RHI should have grown much faster during Exp.~1 than during Exp.~2 (Fig.~\ref{rhi_growth}). 

Comparison of measured debris parameters with RHI growth rates from linear theory suggest that raising the debris beam density would substantially increase the RHI growth rate. Future work will focus on increasing the beam density and raising the  Alfv\'{e}nic Mach number to enter the NRI regime.

\begin{acknowledgments}
This work was supported by the DTRA under Contract No. HDTRA1-12-1-0024, the DOE under Contract Nos. DE-SC0006538:0003 and DE-SC0017900, and by NSF Award No. 1414591.  Thanks to Brent Dane (MIC) for helping to commission the Peening laser system, and Andi Henning for facilitating the loan of the Peening laser from the Space and Naval Warfare Systems Command. Experiments were performed at the UCLA Basic Plasma Science Facility (BaPSF), supported by the DOE under Contract No. DE-FC02-07ER54918  and the NSF under Award No. PHY-1561912. We would like to thank the staff of BaPSF, Z. Lucky, M. Drandell, and T. Ly for their help conducting this experiment. 
\end{acknowledgments}


\begin{thebibliography}{50}%
\makeatletter
\providecommand \@ifxundefined [1]{%
 \@ifx{#1\undefined}
}%
\providecommand \@ifnum [1]{%
 \ifnum #1\expandafter \@firstoftwo
 \else \expandafter \@secondoftwo
 \fi
}%
\providecommand \@ifx [1]{%
 \ifx #1\expandafter \@firstoftwo
 \else \expandafter \@secondoftwo
 \fi
}%
\providecommand \natexlab [1]{#1}%
\providecommand \enquote  [1]{``#1''}%
\providecommand \bibnamefont  [1]{#1}%
\providecommand \bibfnamefont [1]{#1}%
\providecommand \citenamefont [1]{#1}%
\providecommand \href@noop [0]{\@secondoftwo}%
\providecommand \href [0]{\begingroup \@sanitize@url \@href}%
\providecommand \@href[1]{\@@startlink{#1}\@@href}%
\providecommand \@@href[1]{\endgroup#1\@@endlink}%
\providecommand \@sanitize@url [0]{\catcode `\\12\catcode `\$12\catcode
  `\&12\catcode `\#12\catcode `\^12\catcode `\_12\catcode `\%12\relax}%
\providecommand \@@startlink[1]{}%
\providecommand \@@endlink[0]{}%
\providecommand \url  [0]{\begingroup\@sanitize@url \@url }%
\providecommand \@url [1]{\endgroup\@href {#1}{\urlprefix }}%
\providecommand \urlprefix  [0]{URL }%
\providecommand \Eprint [0]{\href }%
\providecommand \doibase [0]{http://dx.doi.org/}%
\providecommand \selectlanguage [0]{\@gobble}%
\providecommand \bibinfo  [0]{\@secondoftwo}%
\providecommand \bibfield  [0]{\@secondoftwo}%
\providecommand \translation [1]{[#1]}%
\providecommand \BibitemOpen [0]{}%
\providecommand \bibitemStop [0]{}%
\providecommand \bibitemNoStop [0]{.\EOS\space}%
\providecommand \EOS [0]{\spacefactor3000\relax}%
\providecommand \BibitemShut  [1]{\csname bibitem#1\endcsname}%
\let\auto@bib@innerbib\@empty
\bibitem [{\citenamefont {Parker}(1961)}]{Parker1961}%
  \BibitemOpen
  \bibfield  {author} {\bibinfo {author} {\bibfnamefont {E.~N.}\ \bibnamefont
  {Parker}},\ }\href@noop {} {\bibfield  {journal} {\bibinfo  {journal}
  {Journal of Nuclear Energy, Part C: Plasma Physics}\ }\textbf {\bibinfo
  {volume} {2}},\ \bibinfo {pages} {146} (\bibinfo {year} {1961})}\BibitemShut
  {NoStop}%
\bibitem [{\citenamefont {Sagdeev}(1966)}]{Sagdeev1966}%
  \BibitemOpen
  \bibfield  {author} {\bibinfo {author} {\bibfnamefont {R.}~\bibnamefont
  {Sagdeev}},\ }\href@noop {} {\bibfield  {journal} {\bibinfo  {journal}
  {Consultants Bureau, New York}\ }\textbf {\bibinfo {volume} {4}},\ \bibinfo
  {pages} {23} (\bibinfo {year} {1966})}\BibitemShut {NoStop}%
\bibitem [{\citenamefont {Treumann}(2009)}]{Treumann2009}%
  \BibitemOpen
  \bibfield  {author} {\bibinfo {author} {\bibfnamefont {R.~A.}\ \bibnamefont
  {Treumann}},\ }\href {\doibase 10.1007/s00159-009-0024-2} {\bibfield
  {journal} {\bibinfo  {journal} {The Astronomy and Astrophysics Review}\
  }\textbf {\bibinfo {volume} {17}},\ \bibinfo {pages} {409} (\bibinfo {year}
  {2009})}\BibitemShut {NoStop}%
\bibitem [{\citenamefont {Mazelle}\ \emph {et~al.}(2003)\citenamefont
  {Mazelle}, \citenamefont {Meziane}, \citenamefont {LeQu{\'e}au},
  \citenamefont {Wilber}, \citenamefont {Eastwood}, \citenamefont {R{\`e}me},
  \citenamefont {Sauvaud}, \citenamefont {Bosqued}, \citenamefont {Dandouras},
  \citenamefont {McCarthy}, \citenamefont {Kistler}, \citenamefont {Klecker},
  \citenamefont {Korth}, \citenamefont {Bavassano-Cattaneo}, \citenamefont
  {Pallocchia}, \citenamefont {Lundin},\ and\ \citenamefont
  {Balogh}}]{Mazelle2003}%
  \BibitemOpen
  \bibfield  {author} {\bibinfo {author} {\bibfnamefont {C.}~\bibnamefont
  {Mazelle}}, \bibinfo {author} {\bibfnamefont {K.}~\bibnamefont {Meziane}},
  \bibinfo {author} {\bibfnamefont {D.}~\bibnamefont {LeQu{\'e}au}}, \bibinfo
  {author} {\bibfnamefont {M.}~\bibnamefont {Wilber}}, \bibinfo {author}
  {\bibfnamefont {J.}~\bibnamefont {Eastwood}}, \bibinfo {author}
  {\bibfnamefont {H.}~\bibnamefont {R{\`e}me}}, \bibinfo {author}
  {\bibfnamefont {J.}~\bibnamefont {Sauvaud}}, \bibinfo {author} {\bibfnamefont
  {J.}~\bibnamefont {Bosqued}}, \bibinfo {author} {\bibfnamefont
  {I.}~\bibnamefont {Dandouras}}, \bibinfo {author} {\bibfnamefont
  {M.}~\bibnamefont {McCarthy}}, \bibinfo {author} {\bibfnamefont
  {L.}~\bibnamefont {Kistler}}, \bibinfo {author} {\bibfnamefont
  {B.}~\bibnamefont {Klecker}}, \bibinfo {author} {\bibfnamefont
  {A.}~\bibnamefont {Korth}}, \bibinfo {author} {\bibfnamefont
  {M.}~\bibnamefont {Bavassano-Cattaneo}}, \bibinfo {author} {\bibfnamefont
  {G.}~\bibnamefont {Pallocchia}}, \bibinfo {author} {\bibfnamefont
  {R.}~\bibnamefont {Lundin}}, \ and\ \bibinfo {author} {\bibfnamefont
  {A.}~\bibnamefont {Balogh}},\ }\href {\doibase
  http://dx.doi.org/10.1016/j.pss.2003.05.002} {\bibfield  {journal} {\bibinfo
  {journal} {Planetary and Space Science}\ }\textbf {\bibinfo {volume} {51}},\
  \bibinfo {pages} {785 } (\bibinfo {year} {2003})}\BibitemShut {NoStop}%
\bibitem [{\citenamefont {Burgess}\ \emph {et~al.}(2005)\citenamefont
  {Burgess}, \citenamefont {Lucek}, \citenamefont {Scholer}, \citenamefont
  {Bale}, \citenamefont {Balikhin}, \citenamefont {Balogh}, \citenamefont
  {Horbury}, \citenamefont {Krasnoselskikh}, \citenamefont {Kucharek},
  \citenamefont {Lemb{\`e}ge}, \citenamefont {M{\"o}bius}, \citenamefont
  {Schwartz}, \citenamefont {Thomsen},\ and\ \citenamefont
  {Walker}}]{Burgess2005quasi}%
  \BibitemOpen
  \bibfield  {author} {\bibinfo {author} {\bibfnamefont {D.}~\bibnamefont
  {Burgess}}, \bibinfo {author} {\bibfnamefont {E.~A.}\ \bibnamefont {Lucek}},
  \bibinfo {author} {\bibfnamefont {M.}~\bibnamefont {Scholer}}, \bibinfo
  {author} {\bibfnamefont {S.~D.}\ \bibnamefont {Bale}}, \bibinfo {author}
  {\bibfnamefont {M.~A.}\ \bibnamefont {Balikhin}}, \bibinfo {author}
  {\bibfnamefont {A.}~\bibnamefont {Balogh}}, \bibinfo {author} {\bibfnamefont
  {T.~S.}\ \bibnamefont {Horbury}}, \bibinfo {author} {\bibfnamefont {V.~V.}\
  \bibnamefont {Krasnoselskikh}}, \bibinfo {author} {\bibfnamefont
  {H.}~\bibnamefont {Kucharek}}, \bibinfo {author} {\bibfnamefont
  {B.}~\bibnamefont {Lemb{\`e}ge}}, \bibinfo {author} {\bibfnamefont
  {E.}~\bibnamefont {M{\"o}bius}}, \bibinfo {author} {\bibfnamefont {S.~J.}\
  \bibnamefont {Schwartz}}, \bibinfo {author} {\bibfnamefont {M.~F.}\
  \bibnamefont {Thomsen}}, \ and\ \bibinfo {author} {\bibfnamefont {S.~N.}\
  \bibnamefont {Walker}},\ }\href {\doibase 10.1007/s11214-005-3832-3}
  {\bibfield  {journal} {\bibinfo  {journal} {Space Science Reviews}\ }\textbf
  {\bibinfo {volume} {118}},\ \bibinfo {pages} {205} (\bibinfo {year}
  {2005})}\BibitemShut {NoStop}%
\bibitem [{\citenamefont {Omidi}\ \emph {et~al.}(1994)\citenamefont {Omidi},
  \citenamefont {Karimabadi}, \citenamefont {Krauss-Varban},\ and\
  \citenamefont {Killen}}]{Omidi1994}%
  \BibitemOpen
  \bibfield  {author} {\bibinfo {author} {\bibfnamefont {N.}~\bibnamefont
  {Omidi}}, \bibinfo {author} {\bibfnamefont {H.}~\bibnamefont {Karimabadi}},
  \bibinfo {author} {\bibfnamefont {D.}~\bibnamefont {Krauss-Varban}}, \ and\
  \bibinfo {author} {\bibfnamefont {K.}~\bibnamefont {Killen}},\ }\enquote
  {\bibinfo {title} {Generation and nonlinear evolution of oblique magnetosonic
  waves: Application to foreshock and comets},}\ in\ \href {\doibase
  10.1029/GM084p0071} {\emph {\bibinfo {booktitle} {Solar System Plasmas in
  Space and Time}}}\ (\bibinfo  {publisher} {American Geophysical Union},\
  \bibinfo {year} {1994})\ pp.\ \bibinfo {pages} {71--84}\BibitemShut {NoStop}%
\bibitem [{\citenamefont {Ostriker}\ and\ \citenamefont
  {McKee}(1988)}]{Ostriker1988}%
  \BibitemOpen
  \bibfield  {author} {\bibinfo {author} {\bibfnamefont {J.~P.}\ \bibnamefont
  {Ostriker}}\ and\ \bibinfo {author} {\bibfnamefont {C.~F.}\ \bibnamefont
  {McKee}},\ }\href@noop {} {\bibfield  {journal} {\bibinfo  {journal} {Rev.
  Mod. Phys.}\ }\textbf {\bibinfo {volume} {60}} (\bibinfo {year}
  {1988})}\BibitemShut {NoStop}%
\bibitem [{\citenamefont {Spicer}\ \emph {et~al.}(1990)\citenamefont {Spicer},
  \citenamefont {Maran},\ and\ \citenamefont {Clark}}]{Spicer1990}%
  \BibitemOpen
  \bibfield  {author} {\bibinfo {author} {\bibfnamefont {D.~S.}\ \bibnamefont
  {Spicer}}, \bibinfo {author} {\bibfnamefont {S.~P.}\ \bibnamefont {Maran}}, \
  and\ \bibinfo {author} {\bibfnamefont {R.}~\bibnamefont {Clark}},\ }\href
  {\doibase 10.1086/168862} {\bibfield  {journal} {\bibinfo  {journal} {The
  Astrophysical Journal}\ }\textbf {\bibinfo {volume} {356}},\ \bibinfo {pages}
  {549} (\bibinfo {year} {1990})}\BibitemShut {NoStop}%
\bibitem [{\citenamefont {Drake}(2000)}]{Drake2000}%
  \BibitemOpen
  \bibfield  {author} {\bibinfo {author} {\bibfnamefont {R.~P.}\ \bibnamefont
  {Drake}},\ }\href@noop {} {\bibfield  {journal} {\bibinfo  {journal} {Physics
  of Plasmas}\ }\textbf {\bibinfo {volume} {7}},\ \bibinfo {pages} {4690}
  (\bibinfo {year} {2000})}\BibitemShut {NoStop}%
\bibitem [{\citenamefont {Thomas}\ \emph {et~al.}(1991)\citenamefont {Thomas},
  \citenamefont {Winske}, \citenamefont {Thomsen},\ and\ \citenamefont
  {Onsager}}]{Thomas1991hybrid}%
  \BibitemOpen
  \bibfield  {author} {\bibinfo {author} {\bibfnamefont {V.~A.}\ \bibnamefont
  {Thomas}}, \bibinfo {author} {\bibfnamefont {D.}~\bibnamefont {Winske}},
  \bibinfo {author} {\bibfnamefont {M.~F.}\ \bibnamefont {Thomsen}}, \ and\
  \bibinfo {author} {\bibfnamefont {T.~G.}\ \bibnamefont {Onsager}},\ }\href
  {http:https://doi.org/10.1029/91JA01092} {\bibfield  {journal} {\bibinfo
  {journal} {Journal of Geophysical Research: Space Physics}\ }\textbf
  {\bibinfo {volume} {96}},\ \bibinfo {pages} {11625} (\bibinfo {year}
  {1991})}\BibitemShut {NoStop}%
\bibitem [{\citenamefont {Parks}\ \emph {et~al.}(2017)\citenamefont {Parks},
  \citenamefont {Lee}, \citenamefont {Fu}, \citenamefont {Lin}, \citenamefont
  {Liu},\ and\ \citenamefont {Yang}}]{Parks2017shocks}%
  \BibitemOpen
  \bibfield  {author} {\bibinfo {author} {\bibfnamefont {G.~K.}\ \bibnamefont
  {Parks}}, \bibinfo {author} {\bibfnamefont {E.}~\bibnamefont {Lee}}, \bibinfo
  {author} {\bibfnamefont {S.~Y.}\ \bibnamefont {Fu}}, \bibinfo {author}
  {\bibfnamefont {N.}~\bibnamefont {Lin}}, \bibinfo {author} {\bibfnamefont
  {Y.}~\bibnamefont {Liu}}, \ and\ \bibinfo {author} {\bibfnamefont {Z.~W.}\
  \bibnamefont {Yang}},\ }\href {\doibase 10.1007/s41614-017-0003-4} {\bibfield
   {journal} {\bibinfo  {journal} {Reviews of Modern Plasma Physics}\ }\textbf
  {\bibinfo {volume} {1}},\ \bibinfo {pages} {1} (\bibinfo {year}
  {2017})}\BibitemShut {NoStop}%
\bibitem [{\citenamefont {Omidi}\ \emph {et~al.}(1990)\citenamefont {Omidi},
  \citenamefont {Quest},\ and\ \citenamefont {Winske}}]{Omidi1990low}%
  \BibitemOpen
  \bibfield  {author} {\bibinfo {author} {\bibfnamefont {N.}~\bibnamefont
  {Omidi}}, \bibinfo {author} {\bibfnamefont {K.~B.}\ \bibnamefont {Quest}}, \
  and\ \bibinfo {author} {\bibfnamefont {D.}~\bibnamefont {Winske}},\ }\href
  {\doibase 10.1029/JA095iA12p20717} {\bibfield  {journal} {\bibinfo  {journal}
  {Journal of Geophysical Research: Space Physics}\ }\textbf {\bibinfo {volume}
  {95}} (\bibinfo {year} {1990}),\ 10.1029/JA095iA12p20717}\BibitemShut
  {NoStop}%
\bibitem [{\citenamefont {Burgess}(1989)}]{Burgess1989cyclic}%
  \BibitemOpen
  \bibfield  {author} {\bibinfo {author} {\bibfnamefont {D.}~\bibnamefont
  {Burgess}},\ }\href {\doibase 10.1029/GL016i005p00345} {\bibfield  {journal}
  {\bibinfo  {journal} {Geophysical Research Letters}\ }\textbf {\bibinfo
  {volume} {16}},\ \bibinfo {pages} {345} (\bibinfo {year} {1989})}\BibitemShut
  {NoStop}%
\bibitem [{\citenamefont {Blandford}\ and\ \citenamefont
  {Eichler}(1987)}]{Blandford1987particle}%
  \BibitemOpen
  \bibfield  {author} {\bibinfo {author} {\bibfnamefont {R.}~\bibnamefont
  {Blandford}}\ and\ \bibinfo {author} {\bibfnamefont {D.}~\bibnamefont
  {Eichler}},\ }\href {\doibase http://dx.doi.org/10.1016/0370-1573(87)90134-7}
  {\bibfield  {journal} {\bibinfo  {journal} {Physics Reports}\ }\textbf
  {\bibinfo {volume} {154}},\ \bibinfo {pages} {1 } (\bibinfo {year}
  {1987})}\BibitemShut {NoStop}%
\bibitem [{\citenamefont {Bell}(2004)}]{Bell2004turbulent}%
  \BibitemOpen
  \bibfield  {author} {\bibinfo {author} {\bibfnamefont {A.~R.}\ \bibnamefont
  {Bell}},\ }\href@noop {} {\bibfield  {journal} {\bibinfo  {journal} {Monthly
  Notices of the Royal Astronomical Society}\ }\textbf {\bibinfo {volume}
  {353}},\ \bibinfo {pages} {550} (\bibinfo {year} {2004})}\BibitemShut
  {NoStop}%
\bibitem [{\citenamefont {Papadopoulos}\ \emph {et~al.}(1987)\citenamefont
  {Papadopoulos}, \citenamefont {Huba},\ and\ \citenamefont
  {Lui}}]{Papadopoulos1987}%
  \BibitemOpen
  \bibfield  {author} {\bibinfo {author} {\bibfnamefont {K.}~\bibnamefont
  {Papadopoulos}}, \bibinfo {author} {\bibfnamefont {J.~D.}\ \bibnamefont
  {Huba}}, \ and\ \bibinfo {author} {\bibfnamefont {A.~T.~Y.}\ \bibnamefont
  {Lui}},\ }\href {\doibase 10.1029/JA092iA01p00047} {\bibfield  {journal}
  {\bibinfo  {journal} {Journal of Geophysical Research: Space Physics}\
  }\textbf {\bibinfo {volume} {92}},\ \bibinfo {pages} {47} (\bibinfo {year}
  {1987})}\BibitemShut {NoStop}%
\bibitem [{\citenamefont {Bondarenko}\ \emph {et~al.}(2017)\citenamefont
  {Bondarenko}, \citenamefont {Schaeffer}, \citenamefont {Everson},
  \citenamefont {Clark}, \citenamefont {Lee}, \citenamefont {Constantin},
  \citenamefont {Vincena}, \citenamefont {Van~Compernolle}, \citenamefont
  {Tripathi}, \citenamefont {Winske},\ and\ \citenamefont
  {Niemann}}]{Bondarenko2017}%
  \BibitemOpen
  \bibfield  {author} {\bibinfo {author} {\bibfnamefont {A.~S.}\ \bibnamefont
  {Bondarenko}}, \bibinfo {author} {\bibfnamefont {D.~B.}\ \bibnamefont
  {Schaeffer}}, \bibinfo {author} {\bibfnamefont {E.~T.}\ \bibnamefont
  {Everson}}, \bibinfo {author} {\bibfnamefont {S.~E.}\ \bibnamefont {Clark}},
  \bibinfo {author} {\bibfnamefont {B.~R.}\ \bibnamefont {Lee}}, \bibinfo
  {author} {\bibfnamefont {C.~G.}\ \bibnamefont {Constantin}}, \bibinfo
  {author} {\bibfnamefont {S.}~\bibnamefont {Vincena}}, \bibinfo {author}
  {\bibfnamefont {B.}~\bibnamefont {Van~Compernolle}}, \bibinfo {author}
  {\bibfnamefont {S.~K.~P.}\ \bibnamefont {Tripathi}}, \bibinfo {author}
  {\bibfnamefont {D.}~\bibnamefont {Winske}}, \ and\ \bibinfo {author}
  {\bibfnamefont {C.}~\bibnamefont {Niemann}},\ }\href
  {http://dx.doi.org/10.1038/nphys4041} {\bibfield  {journal} {\bibinfo
  {journal} {Nat Phys}\ }\textbf {\bibinfo {volume} {13}},\ \bibinfo {pages}
  {573} (\bibinfo {year} {2017})}\BibitemShut {NoStop}%
\bibitem [{\citenamefont {Newbury}\ \emph {et~al.}(1998)\citenamefont
  {Newbury}, \citenamefont {Russell},\ and\ \citenamefont
  {Gedalin}}]{Newbury1998ramp}%
  \BibitemOpen
  \bibfield  {author} {\bibinfo {author} {\bibfnamefont {J.~A.}\ \bibnamefont
  {Newbury}}, \bibinfo {author} {\bibfnamefont {C.~T.}\ \bibnamefont
  {Russell}}, \ and\ \bibinfo {author} {\bibfnamefont {M.}~\bibnamefont
  {Gedalin}},\ }\href {\doibase 10.1029/1998JA900024} {\bibfield  {journal}
  {\bibinfo  {journal} {Journal of Geophysical Research: Space Physics}\
  }\textbf {\bibinfo {volume} {103}} (\bibinfo {year} {1998}),\
  10.1029/1998JA900024}\BibitemShut {NoStop}%
\bibitem [{\citenamefont {Niemann}\ \emph {et~al.}(2014)\citenamefont
  {Niemann}, \citenamefont {Gekelman}, \citenamefont {Constantin},
  \citenamefont {Everson}, \citenamefont {Schaeffer}, \citenamefont
  {Bondarenko}, \citenamefont {Clark}, \citenamefont {Winske}, \citenamefont
  {Vincena}, \citenamefont {Compernolle},\ and\ \citenamefont
  {Pribyl}}]{Niemann2014}%
  \BibitemOpen
  \bibfield  {author} {\bibinfo {author} {\bibfnamefont {C.}~\bibnamefont
  {Niemann}}, \bibinfo {author} {\bibfnamefont {W.}~\bibnamefont {Gekelman}},
  \bibinfo {author} {\bibfnamefont {C.~G.}\ \bibnamefont {Constantin}},
  \bibinfo {author} {\bibfnamefont {E.~T.}\ \bibnamefont {Everson}}, \bibinfo
  {author} {\bibfnamefont {D.~B.}\ \bibnamefont {Schaeffer}}, \bibinfo {author}
  {\bibfnamefont {A.~S.}\ \bibnamefont {Bondarenko}}, \bibinfo {author}
  {\bibfnamefont {S.~E.}\ \bibnamefont {Clark}}, \bibinfo {author}
  {\bibfnamefont {D.}~\bibnamefont {Winske}}, \bibinfo {author} {\bibfnamefont
  {S.}~\bibnamefont {Vincena}}, \bibinfo {author} {\bibfnamefont {B.~V.}\
  \bibnamefont {Compernolle}}, \ and\ \bibinfo {author} {\bibfnamefont
  {P.}~\bibnamefont {Pribyl}},\ }\href@noop {} {\bibfield  {journal} {\bibinfo
  {journal} {Geophysical Research Letters}\ }\textbf {\bibinfo {volume} {41}},\
  \bibinfo {pages} {7413} (\bibinfo {year} {2014})}\BibitemShut {NoStop}%
\bibitem [{\citenamefont {Schaeffer}\ \emph
  {et~al.}(2017{\natexlab{a}})\citenamefont {Schaeffer}, \citenamefont
  {Winske}, \citenamefont {Larson}, \citenamefont {Cowee}, \citenamefont
  {Constantin}, \citenamefont {Bondarenko}, \citenamefont {Clark},\ and\
  \citenamefont {Niemann}}]{Schaeffer2017}%
  \BibitemOpen
  \bibfield  {author} {\bibinfo {author} {\bibfnamefont {D.~B.}\ \bibnamefont
  {Schaeffer}}, \bibinfo {author} {\bibfnamefont {D.}~\bibnamefont {Winske}},
  \bibinfo {author} {\bibfnamefont {D.~J.}\ \bibnamefont {Larson}}, \bibinfo
  {author} {\bibfnamefont {M.~M.}\ \bibnamefont {Cowee}}, \bibinfo {author}
  {\bibfnamefont {C.~G.}\ \bibnamefont {Constantin}}, \bibinfo {author}
  {\bibfnamefont {A.~S.}\ \bibnamefont {Bondarenko}}, \bibinfo {author}
  {\bibfnamefont {S.~E.}\ \bibnamefont {Clark}}, \ and\ \bibinfo {author}
  {\bibfnamefont {C.}~\bibnamefont {Niemann}},\ }\href {\doibase
  10.1063/1.4978882} {\bibfield  {journal} {\bibinfo  {journal} {Physics of
  Plasmas}\ }\textbf {\bibinfo {volume} {24}} (\bibinfo {year}
  {2017}{\natexlab{a}}),\ 10.1063/1.4978882}\BibitemShut {NoStop}%
\bibitem [{\citenamefont {Schaeffer}\ \emph
  {et~al.}(2017{\natexlab{b}})\citenamefont {Schaeffer}, \citenamefont {Fox},
  \citenamefont {Haberberger}, \citenamefont {Fiksel}, \citenamefont
  {Bhattacharjee}, \citenamefont {Barnak}, \citenamefont {Hu},\ and\
  \citenamefont {Germaschewski}}]{Schaeffer2017generation}%
  \BibitemOpen
  \bibfield  {author} {\bibinfo {author} {\bibfnamefont {D.~B.}\ \bibnamefont
  {Schaeffer}}, \bibinfo {author} {\bibfnamefont {W.}~\bibnamefont {Fox}},
  \bibinfo {author} {\bibfnamefont {D.}~\bibnamefont {Haberberger}}, \bibinfo
  {author} {\bibfnamefont {G.}~\bibnamefont {Fiksel}}, \bibinfo {author}
  {\bibfnamefont {A.}~\bibnamefont {Bhattacharjee}}, \bibinfo {author}
  {\bibfnamefont {D.~H.}\ \bibnamefont {Barnak}}, \bibinfo {author}
  {\bibfnamefont {S.~X.}\ \bibnamefont {Hu}}, \ and\ \bibinfo {author}
  {\bibfnamefont {K.}~\bibnamefont {Germaschewski}},\ }\href@noop {} {\bibfield
   {journal} {\bibinfo  {journal} {Phys. Rev. Lett.}\ }\textbf {\bibinfo
  {volume} {119}} (\bibinfo {year} {2017}{\natexlab{b}})}\BibitemShut {NoStop}%
\bibitem [{\citenamefont {Winske}\ and\ \citenamefont
  {Leroy}(1984)}]{Winske1984}%
  \BibitemOpen
  \bibfield  {author} {\bibinfo {author} {\bibfnamefont {D.}~\bibnamefont
  {Winske}}\ and\ \bibinfo {author} {\bibfnamefont {M.~M.}\ \bibnamefont
  {Leroy}},\ }\href@noop {} {\bibfield  {journal} {\bibinfo  {journal} {Journal
  of Geophysical Research}\ }\textbf {\bibinfo {volume} {89}},\ \bibinfo
  {pages} {2673} (\bibinfo {year} {1984})}\BibitemShut {NoStop}%
\bibitem [{\citenamefont {Gary}(1991)}]{Gary1991}%
  \BibitemOpen
  \bibfield  {author} {\bibinfo {author} {\bibfnamefont {S.~P.}\ \bibnamefont
  {Gary}},\ }\href {\doibase 10.1007/BF00196632} {\bibfield  {journal}
  {\bibinfo  {journal} {Space Science Reviews}\ }\textbf {\bibinfo {volume}
  {56}},\ \bibinfo {pages} {373} (\bibinfo {year} {1991})}\BibitemShut
  {NoStop}%
\bibitem [{\citenamefont {Gary}(1985)}]{Gary1985}%
  \BibitemOpen
  \bibfield  {author} {\bibinfo {author} {\bibfnamefont {S.~P.}\ \bibnamefont
  {Gary}},\ }\href@noop {} {\bibfield  {journal} {\bibinfo  {journal} {The
  Astrophysical Journal}\ }\textbf {\bibinfo {volume} {288}},\ \bibinfo {pages}
  {342} (\bibinfo {year} {1985})}\BibitemShut {NoStop}%
\bibitem [{\citenamefont {Winske}\ \emph {et~al.}(1985)\citenamefont {Winske},
  \citenamefont {Wu}, \citenamefont {Li}, \citenamefont {Mou},\ and\
  \citenamefont {Guo}}]{Winske1985}%
  \BibitemOpen
  \bibfield  {author} {\bibinfo {author} {\bibfnamefont {D.}~\bibnamefont
  {Winske}}, \bibinfo {author} {\bibfnamefont {C.~S.}\ \bibnamefont {Wu}},
  \bibinfo {author} {\bibfnamefont {Y.~Y.}\ \bibnamefont {Li}}, \bibinfo
  {author} {\bibfnamefont {Z.~Z.}\ \bibnamefont {Mou}}, \ and\ \bibinfo
  {author} {\bibfnamefont {S.~Y.}\ \bibnamefont {Guo}},\ }\href {\doibase
  10.1029/JA090iA03p02713} {\bibfield  {journal} {\bibinfo  {journal} {Journal
  of Geophysical Research: Space Physics}\ }\textbf {\bibinfo {volume} {90}},\
  \bibinfo {pages} {2713} (\bibinfo {year} {1985})}\BibitemShut {NoStop}%
\bibitem [{\citenamefont {Burgess}(1995)}]{Burgess1995}%
  \BibitemOpen
  \bibfield  {author} {\bibinfo {author} {\bibfnamefont {D.}~\bibnamefont
  {Burgess}},\ }\href@noop {} {\bibfield  {journal} {\bibinfo  {journal}
  {Advanced Space Research}\ }\textbf {\bibinfo {volume} {15}} (\bibinfo {year}
  {1995})}\BibitemShut {NoStop}%
\bibitem [{\citenamefont {Weidl}\ \emph {et~al.}(2016)\citenamefont {Weidl},
  \citenamefont {Winske}, \citenamefont {Jenko},\ and\ \citenamefont
  {Niemann}}]{Weidl2016}%
  \BibitemOpen
  \bibfield  {author} {\bibinfo {author} {\bibfnamefont {M.~S.}\ \bibnamefont
  {Weidl}}, \bibinfo {author} {\bibfnamefont {D.}~\bibnamefont {Winske}},
  \bibinfo {author} {\bibfnamefont {F.}~\bibnamefont {Jenko}}, \ and\ \bibinfo
  {author} {\bibfnamefont {C.}~\bibnamefont {Niemann}},\ }\href@noop {}
  {\bibfield  {journal} {\bibinfo  {journal} {Physics of Plasmas}\ }\textbf
  {\bibinfo {volume} {23}} (\bibinfo {year} {2016})}\BibitemShut {NoStop}%
\bibitem [{\citenamefont {Gekelman}\ \emph {et~al.}(2016)\citenamefont
  {Gekelman}, \citenamefont {Pribyl}, \citenamefont {Lucky}, \citenamefont
  {Drandell}, \citenamefont {Leneman}, \citenamefont {Maggs}, \citenamefont
  {Vincena}, \citenamefont {Van~Compernolle}, \citenamefont {Tripathi},
  \citenamefont {Morales}, \citenamefont {Carter}, \citenamefont {Wang},\ and\
  \citenamefont {DeHaas}}]{Gekelman2016}%
  \BibitemOpen
  \bibfield  {author} {\bibinfo {author} {\bibfnamefont {W.}~\bibnamefont
  {Gekelman}}, \bibinfo {author} {\bibfnamefont {P.}~\bibnamefont {Pribyl}},
  \bibinfo {author} {\bibfnamefont {Z.}~\bibnamefont {Lucky}}, \bibinfo
  {author} {\bibfnamefont {M.}~\bibnamefont {Drandell}}, \bibinfo {author}
  {\bibfnamefont {D.}~\bibnamefont {Leneman}}, \bibinfo {author} {\bibfnamefont
  {J.}~\bibnamefont {Maggs}}, \bibinfo {author} {\bibfnamefont
  {S.}~\bibnamefont {Vincena}}, \bibinfo {author} {\bibfnamefont
  {B.}~\bibnamefont {Van~Compernolle}}, \bibinfo {author} {\bibfnamefont
  {S.~K.~P.}\ \bibnamefont {Tripathi}}, \bibinfo {author} {\bibfnamefont
  {G.}~\bibnamefont {Morales}}, \bibinfo {author} {\bibfnamefont {T.~A.}\
  \bibnamefont {Carter}}, \bibinfo {author} {\bibfnamefont {Y.}~\bibnamefont
  {Wang}}, \ and\ \bibinfo {author} {\bibfnamefont {T.}~\bibnamefont
  {DeHaas}},\ }\href@noop {} {\bibfield  {journal} {\bibinfo  {journal} {Review
  of Scientific Instruments}\ }\textbf {\bibinfo {volume} {87}} (\bibinfo
  {year} {2016})}\BibitemShut {NoStop}%
\bibitem [{\citenamefont {Niemann}\ \emph {et~al.}(2012)\citenamefont
  {Niemann}, \citenamefont {Constantin}, \citenamefont {Schaeffer},
  \citenamefont {Tauschwitz}, \citenamefont {Weiland}, \citenamefont {Lucky},
  \citenamefont {Gekelman}, \citenamefont {Everson},\ and\ \citenamefont
  {Winske}}]{Niemann2012}%
  \BibitemOpen
  \bibfield  {author} {\bibinfo {author} {\bibfnamefont {C.}~\bibnamefont
  {Niemann}}, \bibinfo {author} {\bibfnamefont {C.~G.}\ \bibnamefont
  {Constantin}}, \bibinfo {author} {\bibfnamefont {D.~B.}\ \bibnamefont
  {Schaeffer}}, \bibinfo {author} {\bibfnamefont {A.}~\bibnamefont
  {Tauschwitz}}, \bibinfo {author} {\bibfnamefont {T.}~\bibnamefont {Weiland}},
  \bibinfo {author} {\bibfnamefont {Z.}~\bibnamefont {Lucky}}, \bibinfo
  {author} {\bibfnamefont {W.}~\bibnamefont {Gekelman}}, \bibinfo {author}
  {\bibfnamefont {E.~T.}\ \bibnamefont {Everson}}, \ and\ \bibinfo {author}
  {\bibfnamefont {D.}~\bibnamefont {Winske}},\ }\href@noop {} {\bibfield
  {journal} {\bibinfo  {journal} {Journal of Instrumentation}\ }\textbf
  {\bibinfo {volume} {7}} (\bibinfo {year} {2012})}\BibitemShut {NoStop}%
\bibitem [{\citenamefont {Gary}\ and\ \citenamefont
  {Feldman}(1978)}]{Gary1978}%
  \BibitemOpen
  \bibfield  {author} {\bibinfo {author} {\bibfnamefont {S.~P.}\ \bibnamefont
  {Gary}}\ and\ \bibinfo {author} {\bibfnamefont {W.~C.}\ \bibnamefont
  {Feldman}},\ }\href {\doibase http://dx.doi.org/10.1063/1.862081} {\bibfield
  {journal} {\bibinfo  {journal} {Physics of Fluids}\ }\textbf {\bibinfo
  {volume} {21}},\ \bibinfo {pages} {72} (\bibinfo {year} {1978})}\BibitemShut
  {NoStop}%
\bibitem [{\citenamefont {Gary}\ \emph {et~al.}(1984)\citenamefont {Gary},
  \citenamefont {Smith}, \citenamefont {Lee}, \citenamefont {Goldstein},\ and\
  \citenamefont {Forslund}}]{Gary1984}%
  \BibitemOpen
  \bibfield  {author} {\bibinfo {author} {\bibfnamefont {S.~P.}\ \bibnamefont
  {Gary}}, \bibinfo {author} {\bibfnamefont {C.~W.}\ \bibnamefont {Smith}},
  \bibinfo {author} {\bibfnamefont {M.~A.}\ \bibnamefont {Lee}}, \bibinfo
  {author} {\bibfnamefont {M.~L.}\ \bibnamefont {Goldstein}}, \ and\ \bibinfo
  {author} {\bibfnamefont {D.~W.}\ \bibnamefont {Forslund}},\ }\href {\doibase
  http://dx.doi.org/10.1063/1.864797} {\bibfield  {journal} {\bibinfo
  {journal} {Physics of Fluids}\ }\textbf {\bibinfo {volume} {27}},\ \bibinfo
  {pages} {1852} (\bibinfo {year} {1984})}\BibitemShut {NoStop}%
\bibitem [{\citenamefont {Weidl}\ \emph {et~al.}(2017)\citenamefont {Weidl},
  \citenamefont {Heuer}, \citenamefont {Schaeffer}, \citenamefont {Dorst},
  \citenamefont {Winske}, \citenamefont {Constantin},\ and\ \citenamefont
  {Niemann}}]{Weidl2017towards}%
  \BibitemOpen
  \bibfield  {author} {\bibinfo {author} {\bibfnamefont {M.~S.}\ \bibnamefont
  {Weidl}}, \bibinfo {author} {\bibfnamefont {P.}~\bibnamefont {Heuer}},
  \bibinfo {author} {\bibfnamefont {D.}~\bibnamefont {Schaeffer}}, \bibinfo
  {author} {\bibfnamefont {R.}~\bibnamefont {Dorst}}, \bibinfo {author}
  {\bibfnamefont {D.}~\bibnamefont {Winske}}, \bibinfo {author} {\bibfnamefont
  {C.}~\bibnamefont {Constantin}}, \ and\ \bibinfo {author} {\bibfnamefont
  {C.}~\bibnamefont {Niemann}},\ }\href
  {http://stacks.iop.org/1742-6596/900/i=1/a=012020} {\bibfield  {journal}
  {\bibinfo  {journal} {Journal of Physics: Conference Series}\ }\textbf
  {\bibinfo {volume} {900}},\ \bibinfo {pages} {012020} (\bibinfo {year}
  {2017})}\BibitemShut {NoStop}%
\bibitem [{\citenamefont {Winske}\ and\ \citenamefont
  {Gary}(1986)}]{Winske1986}%
  \BibitemOpen
  \bibfield  {author} {\bibinfo {author} {\bibfnamefont {D.}~\bibnamefont
  {Winske}}\ and\ \bibinfo {author} {\bibfnamefont {S.~P.}\ \bibnamefont
  {Gary}},\ }\href {\doibase 10.1029/JA091iA06p06825} {\bibfield  {journal}
  {\bibinfo  {journal} {Journal of Geophysical Research: Space Physics}\
  }\textbf {\bibinfo {volume} {91}},\ \bibinfo {pages} {6825} (\bibinfo {year}
  {1986})}\BibitemShut {NoStop}%
\bibitem [{\citenamefont {Hackel}\ \emph {et~al.}(1993)\citenamefont {Hackel},
  \citenamefont {Miller},\ and\ \citenamefont {Dane}}]{Hackel1993}%
  \BibitemOpen
  \bibfield  {author} {\bibinfo {author} {\bibfnamefont {L.}~\bibnamefont
  {Hackel}}, \bibinfo {author} {\bibfnamefont {J.}~\bibnamefont {Miller}}, \
  and\ \bibinfo {author} {\bibfnamefont {C.}~\bibnamefont {Dane}},\ }\href@noop
  {} {\bibfield  {journal} {\bibinfo  {journal} {International Journal of
  Nonlinear Optical Physics}\ }\textbf {\bibinfo {volume} {2}},\ \bibinfo
  {pages} {171} (\bibinfo {year} {1993})}\BibitemShut {NoStop}%
\bibitem [{\citenamefont {Grun}\ \emph {et~al.}(1981)\citenamefont {Grun},
  \citenamefont {Decoste}, \citenamefont {Ripin},\ and\ \citenamefont
  {Gardner}}]{Grun1981characteristics}%
  \BibitemOpen
  \bibfield  {author} {\bibinfo {author} {\bibfnamefont {J.}~\bibnamefont
  {Grun}}, \bibinfo {author} {\bibfnamefont {R.}~\bibnamefont {Decoste}},
  \bibinfo {author} {\bibfnamefont {B.~H.}\ \bibnamefont {Ripin}}, \ and\
  \bibinfo {author} {\bibfnamefont {J.}~\bibnamefont {Gardner}},\ }\href@noop
  {} {\bibfield  {journal} {\bibinfo  {journal} {Applied Physics Letters}\
  }\textbf {\bibinfo {volume} {39}},\ \bibinfo {pages} {545} (\bibinfo {year}
  {1981})}\BibitemShut {NoStop}%
\bibitem [{\citenamefont {Schaeffer}\ \emph {et~al.}(2016)\citenamefont
  {Schaeffer}, \citenamefont {Bondarenko}, \citenamefont {Everson},
  \citenamefont {Clark}, \citenamefont {Constantin},\ and\ \citenamefont
  {Niemann}}]{Schaeffer2016}%
  \BibitemOpen
  \bibfield  {author} {\bibinfo {author} {\bibfnamefont {D.~B.}\ \bibnamefont
  {Schaeffer}}, \bibinfo {author} {\bibfnamefont {A.~S.}\ \bibnamefont
  {Bondarenko}}, \bibinfo {author} {\bibfnamefont {E.~T.}\ \bibnamefont
  {Everson}}, \bibinfo {author} {\bibfnamefont {S.~E.}\ \bibnamefont {Clark}},
  \bibinfo {author} {\bibfnamefont {C.~G.}\ \bibnamefont {Constantin}}, \ and\
  \bibinfo {author} {\bibfnamefont {C.}~\bibnamefont {Niemann}},\ }\href@noop
  {} {\bibfield  {journal} {\bibinfo  {journal} {Journal of Applied Physics}\
  }\textbf {\bibinfo {volume} {120}},\ \bibinfo {eid} {043301} (\bibinfo {year}
  {2016})}\BibitemShut {NoStop}%
\bibitem [{\citenamefont {Everson}\ \emph {et~al.}(2009)\citenamefont
  {Everson}, \citenamefont {Pribyl}, \citenamefont {Constantin}, \citenamefont
  {Zylstra}, \citenamefont {Schaeffer}, \citenamefont {Kugland},\ and\
  \citenamefont {Niemann}}]{Everson2009}%
  \BibitemOpen
  \bibfield  {author} {\bibinfo {author} {\bibfnamefont {E.~T.}\ \bibnamefont
  {Everson}}, \bibinfo {author} {\bibfnamefont {P.}~\bibnamefont {Pribyl}},
  \bibinfo {author} {\bibfnamefont {C.~G.}\ \bibnamefont {Constantin}},
  \bibinfo {author} {\bibfnamefont {A.}~\bibnamefont {Zylstra}}, \bibinfo
  {author} {\bibfnamefont {D.}~\bibnamefont {Schaeffer}}, \bibinfo {author}
  {\bibfnamefont {N.~L.}\ \bibnamefont {Kugland}}, \ and\ \bibinfo {author}
  {\bibfnamefont {C.}~\bibnamefont {Niemann}},\ }\href@noop {} {\bibfield
  {journal} {\bibinfo  {journal} {Rev. Sci. Insturm.}\ }\textbf {\bibinfo
  {volume} {11}} (\bibinfo {year} {2009})}\BibitemShut {NoStop}%
\bibitem [{\citenamefont {Niemann}\ \emph {et~al.}(2013)\citenamefont
  {Niemann}, \citenamefont {Gekelman}, \citenamefont {Constantin},
  \citenamefont {Everson}, \citenamefont {Schaeffer}, \citenamefont {Clark},
  \citenamefont {Winske}, \citenamefont {Zylstra}, \citenamefont {Pribyl},
  \citenamefont {Tripathi}, \citenamefont {Larson}, \citenamefont {Glenzer},\
  and\ \citenamefont {Bondarenko}}]{Niemann2013}%
  \BibitemOpen
  \bibfield  {author} {\bibinfo {author} {\bibfnamefont {C.}~\bibnamefont
  {Niemann}}, \bibinfo {author} {\bibfnamefont {W.}~\bibnamefont {Gekelman}},
  \bibinfo {author} {\bibfnamefont {C.~G.}\ \bibnamefont {Constantin}},
  \bibinfo {author} {\bibfnamefont {E.~T.}\ \bibnamefont {Everson}}, \bibinfo
  {author} {\bibfnamefont {D.~B.}\ \bibnamefont {Schaeffer}}, \bibinfo {author}
  {\bibfnamefont {S.~E.}\ \bibnamefont {Clark}}, \bibinfo {author}
  {\bibfnamefont {D.}~\bibnamefont {Winske}}, \bibinfo {author} {\bibfnamefont
  {A.~B.}\ \bibnamefont {Zylstra}}, \bibinfo {author} {\bibfnamefont
  {P.}~\bibnamefont {Pribyl}}, \bibinfo {author} {\bibfnamefont {S.~K.~P.}\
  \bibnamefont {Tripathi}}, \bibinfo {author} {\bibfnamefont {D.}~\bibnamefont
  {Larson}}, \bibinfo {author} {\bibfnamefont {S.~H.}\ \bibnamefont {Glenzer}},
  \ and\ \bibinfo {author} {\bibfnamefont {A.~S.}\ \bibnamefont {Bondarenko}},\
  }\href@noop {} {\bibfield  {journal} {\bibinfo  {journal} {Physics of
  Plasmas}\ }\textbf {\bibinfo {volume} {20}} (\bibinfo {year}
  {2013})}\BibitemShut {NoStop}%
\bibitem [{\citenamefont {Heuer}\ \emph {et~al.}(2017)\citenamefont {Heuer},
  \citenamefont {Schaeffer}, \citenamefont {Knall}, \citenamefont {Constantin},
  \citenamefont {Hofer}, \citenamefont {Vincena}, \citenamefont {Tripathi},\
  and\ \citenamefont {Niemann}}]{Heuer2016}%
  \BibitemOpen
  \bibfield  {author} {\bibinfo {author} {\bibfnamefont {P.}~\bibnamefont
  {Heuer}}, \bibinfo {author} {\bibfnamefont {D.}~\bibnamefont {Schaeffer}},
  \bibinfo {author} {\bibfnamefont {E.}~\bibnamefont {Knall}}, \bibinfo
  {author} {\bibfnamefont {C.}~\bibnamefont {Constantin}}, \bibinfo {author}
  {\bibfnamefont {L.}~\bibnamefont {Hofer}}, \bibinfo {author} {\bibfnamefont
  {S.}~\bibnamefont {Vincena}}, \bibinfo {author} {\bibfnamefont
  {S.}~\bibnamefont {Tripathi}}, \ and\ \bibinfo {author} {\bibfnamefont
  {C.}~\bibnamefont {Niemann}},\ }\href {\doibase
  http://dx.doi.org/10.1016/j.hedp.2016.12.003} {\bibfield  {journal} {\bibinfo
   {journal} {High Energy Density Physics}\ }\textbf {\bibinfo {volume} {22}},\
  \bibinfo {pages} {17 } (\bibinfo {year} {2017})}\BibitemShut {NoStop}%
\bibitem [{\citenamefont {Meyer}\ and\ \citenamefont
  {Thiell}(1984)}]{Meyer1984experimental}%
  \BibitemOpen
  \bibfield  {author} {\bibinfo {author} {\bibfnamefont {B.}~\bibnamefont
  {Meyer}}\ and\ \bibinfo {author} {\bibfnamefont {G.}~\bibnamefont {Thiell}},\
  }\href@noop {} {\bibfield  {journal} {\bibinfo  {journal} {The Physics of
  Fluids}\ }\textbf {\bibinfo {volume} {27}},\ \bibinfo {pages} {302} (\bibinfo
  {year} {1984})}\BibitemShut {NoStop}%
\bibitem [{\citenamefont {Rumsby}\ and\ \citenamefont
  {Paul}(1974)}]{Rumsby1974}%
  \BibitemOpen
  \bibfield  {author} {\bibinfo {author} {\bibfnamefont {P.~T.}\ \bibnamefont
  {Rumsby}}\ and\ \bibinfo {author} {\bibfnamefont {J.~W.~M.}\ \bibnamefont
  {Paul}},\ }\href {http://stacks.iop.org/0032-1028/16/i=3/a=002} {\bibfield
  {journal} {\bibinfo  {journal} {Plasma Physics}\ }\textbf {\bibinfo {volume}
  {16}},\ \bibinfo {pages} {247} (\bibinfo {year} {1974})}\BibitemShut
  {NoStop}%
\bibitem [{\citenamefont {Nahar}\ and\ \citenamefont
  {Pradhan}(1997)}]{Nahar1997}%
  \BibitemOpen
  \bibfield  {author} {\bibinfo {author} {\bibfnamefont {S.~N.}\ \bibnamefont
  {Nahar}}\ and\ \bibinfo {author} {\bibfnamefont {A.~K.}\ \bibnamefont
  {Pradhan}},\ }\href@noop {} {\bibfield  {journal} {\bibinfo  {journal} {The
  Astrophysical Journal Supplement Series}\ }\textbf {\bibinfo {volume}
  {111}},\ \bibinfo {pages} {339} (\bibinfo {year} {1997})}\BibitemShut
  {NoStop}%
\bibitem [{\citenamefont {Dijkkamp}\ \emph {et~al.}(1985)\citenamefont
  {Dijkkamp}, \citenamefont {Ciric}, \citenamefont {Vileg}, \citenamefont
  {de~Boer},\ and\ \citenamefont {de~Heer}}]{Dijkkamp1985}%
  \BibitemOpen
  \bibfield  {author} {\bibinfo {author} {\bibfnamefont {D.}~\bibnamefont
  {Dijkkamp}}, \bibinfo {author} {\bibfnamefont {D.}~\bibnamefont {Ciric}},
  \bibinfo {author} {\bibfnamefont {E.}~\bibnamefont {Vileg}}, \bibinfo
  {author} {\bibfnamefont {A.}~\bibnamefont {de~Boer}}, \ and\ \bibinfo
  {author} {\bibfnamefont {F.~J.}\ \bibnamefont {de~Heer}},\ }\href
  {http://stacks.iop.org/0022-3700/18/i=24/a=017} {\bibfield  {journal}
  {\bibinfo  {journal} {Journal of Physics B: Atomic and Molecular Physics}\
  }\textbf {\bibinfo {volume} {18}},\ \bibinfo {pages} {4763} (\bibinfo {year}
  {1985})}\BibitemShut {NoStop}%
\bibitem [{\citenamefont {Huba}(2017)}]{NRLformulary}%
  \BibitemOpen
  \bibfield  {author} {\bibinfo {author} {\bibfnamefont {J.}~\bibnamefont
  {Huba}},\ }\href@noop {} {\emph {\bibinfo {title} {NRL Plasma Formulary}}}\
  (\bibinfo  {publisher} {Naval Research Laboratory},\ \bibinfo {year}
  {2017})\BibitemShut {NoStop}%
\bibitem [{\citenamefont {Terasawa}\ \emph {et~al.}(1986)\citenamefont
  {Terasawa}, \citenamefont {Hoshino}, \citenamefont {Sakai},\ and\
  \citenamefont {Hada}}]{Terasawa1986decay}%
  \BibitemOpen
  \bibfield  {author} {\bibinfo {author} {\bibfnamefont {T.}~\bibnamefont
  {Terasawa}}, \bibinfo {author} {\bibfnamefont {M.}~\bibnamefont {Hoshino}},
  \bibinfo {author} {\bibfnamefont {J.}~\bibnamefont {Sakai}}, \ and\ \bibinfo
  {author} {\bibfnamefont {T.}~\bibnamefont {Hada}},\ }\href {\doibase
  10.1029/JA091iA04p04171} {\bibfield  {journal} {\bibinfo  {journal} {Journal
  of Geophysical Research}\ }\textbf {\bibinfo {volume} {91}},\ \bibinfo
  {pages} {4171} (\bibinfo {year} {1986})}\BibitemShut {NoStop}%
\bibitem [{\citenamefont {Vincena}\ \emph {et~al.}(2008)\citenamefont
  {Vincena}, \citenamefont {Gekelman}, \citenamefont {Zeeland}, \citenamefont
  {Maggs},\ and\ \citenamefont {Collette}}]{Vincena2008quasielectrostatic}%
  \BibitemOpen
  \bibfield  {author} {\bibinfo {author} {\bibfnamefont {S.}~\bibnamefont
  {Vincena}}, \bibinfo {author} {\bibfnamefont {W.}~\bibnamefont {Gekelman}},
  \bibinfo {author} {\bibfnamefont {M.~A.~V.}\ \bibnamefont {Zeeland}},
  \bibinfo {author} {\bibfnamefont {J.}~\bibnamefont {Maggs}}, \ and\ \bibinfo
  {author} {\bibfnamefont {A.}~\bibnamefont {Collette}},\ }\href@noop {}
  {\bibfield  {journal} {\bibinfo  {journal} {Physics of Plasmas}\ }\textbf
  {\bibinfo {volume} {15}} (\bibinfo {year} {2008})}\BibitemShut {NoStop}%
\bibitem [{\citenamefont {VanZeeland}\ \emph {et~al.}(2001)\citenamefont
  {VanZeeland}, \citenamefont {Gekelman}, \citenamefont {Vincena},\ and\
  \citenamefont {Dimonte}}]{VanZeeland2001}%
  \BibitemOpen
  \bibfield  {author} {\bibinfo {author} {\bibfnamefont {M.}~\bibnamefont
  {VanZeeland}}, \bibinfo {author} {\bibfnamefont {W.}~\bibnamefont
  {Gekelman}}, \bibinfo {author} {\bibfnamefont {S.}~\bibnamefont {Vincena}}, \
  and\ \bibinfo {author} {\bibfnamefont {G.}~\bibnamefont {Dimonte}},\
  }\href@noop {} {\bibfield  {journal} {\bibinfo  {journal} {Phys. Rev. Lett.}\
  }\textbf {\bibinfo {volume} {87}} (\bibinfo {year} {2001})}\BibitemShut
  {NoStop}%
\bibitem [{\citenamefont {Van~Compernolle}\ \emph {et~al.}(2008)\citenamefont
  {Van~Compernolle}, \citenamefont {Morales},\ and\ \citenamefont
  {Gekelman}}]{VanCompernolle2008}%
  \BibitemOpen
  \bibfield  {author} {\bibinfo {author} {\bibfnamefont {B.}~\bibnamefont
  {Van~Compernolle}}, \bibinfo {author} {\bibfnamefont {G.~J.}\ \bibnamefont
  {Morales}}, \ and\ \bibinfo {author} {\bibfnamefont {W.}~\bibnamefont
  {Gekelman}},\ }\href@noop {} {\bibfield  {journal} {\bibinfo  {journal}
  {Physics of Plasmas}\ }\textbf {\bibinfo {volume} {15}} (\bibinfo {year}
  {2008})}\BibitemShut {NoStop}%
\bibitem [{\citenamefont {Gekelman}\ \emph {et~al.}(2011)\citenamefont
  {Gekelman}, \citenamefont {Vincena}, \citenamefont {Compernolle},
  \citenamefont {Morales}, \citenamefont {Maggs}, \citenamefont {Pribyl},\ and\
  \citenamefont {Carter}}]{Gekelman2011}%
  \BibitemOpen
  \bibfield  {author} {\bibinfo {author} {\bibfnamefont {W.}~\bibnamefont
  {Gekelman}}, \bibinfo {author} {\bibfnamefont {S.}~\bibnamefont {Vincena}},
  \bibinfo {author} {\bibfnamefont {B.~V.}\ \bibnamefont {Compernolle}},
  \bibinfo {author} {\bibfnamefont {G.~J.}\ \bibnamefont {Morales}}, \bibinfo
  {author} {\bibfnamefont {J.~E.}\ \bibnamefont {Maggs}}, \bibinfo {author}
  {\bibfnamefont {P.}~\bibnamefont {Pribyl}}, \ and\ \bibinfo {author}
  {\bibfnamefont {T.~A.}\ \bibnamefont {Carter}},\ }\href@noop {} {\bibfield
  {journal} {\bibinfo  {journal} {Physics of Plasmas}\ }\textbf {\bibinfo
  {volume} {18}} (\bibinfo {year} {2011})}\BibitemShut {NoStop}%
\bibitem [{\citenamefont {Tripathi}\ \emph {et~al.}(2015)\citenamefont
  {Tripathi}, \citenamefont {Van~Compernolle}, \citenamefont {Gekelman},
  \citenamefont {Pribyl},\ and\ \citenamefont {Heidbrink}}]{Tripathi2015}%
  \BibitemOpen
  \bibfield  {author} {\bibinfo {author} {\bibfnamefont {S.~K.~P.}\
  \bibnamefont {Tripathi}}, \bibinfo {author} {\bibfnamefont {B.}~\bibnamefont
  {Van~Compernolle}}, \bibinfo {author} {\bibfnamefont {W.}~\bibnamefont
  {Gekelman}}, \bibinfo {author} {\bibfnamefont {P.}~\bibnamefont {Pribyl}}, \
  and\ \bibinfo {author} {\bibfnamefont {W.}~\bibnamefont {Heidbrink}},\
  }\href@noop {} {\bibfield  {journal} {\bibinfo  {journal} {Phys. Rev. E}\
  }\textbf {\bibinfo {volume} {91}} (\bibinfo {year} {2015})}\BibitemShut
  {NoStop}%
\end{thebibliography}


\end{document}